\documentclass[journal,comsoc]{IEEEtran}
\usepackage[T1]{fontenc}
\usepackage{amsmath}
\usepackage[cmintegrals]{newtxmath}
\usepackage{url}
\usepackage{soul,color,graphicx,float,epstopdf}
\usepackage{tablefootnote,textcomp,gensymb}
\usepackage{eurosym,booktabs,multirow}
\usepackage[caption=false]{subfig}
\usepackage{amsfonts} 
\usepackage{algorithmic}
\usepackage{array}
\usepackage{stfloats}
\usepackage{float}
\usepackage{mathtools}
\usepackage{graphicx}
\usepackage{pdfpages}
\usepackage{subfig} 
\usepackage{balance}
\usepackage{xcolor}
\usepackage{comment}
\usepackage[normalem]{ulem}
\usepackage{makecell}
\usepackage[implicit=false]{hyperref}
\usepackage{tikz}
\usetikzlibrary{arrows,shapes,positioning,shadows,trees}

\tikzset{
  basic/.style  = {draw, text width=4cm, font=\sffamily, rectangle},
  root/.style   = {basic, rounded corners=2pt, thin, align=center,
                   fill=gray!20},
  level 2/.style = {basic, rounded corners=6pt, thin,align=center, fill=blue!10,
                   text width=9em},
  level 3/.style = {basic, thin, align=left, fill=yellow!5, text width=10em}
}

\definecolor{Orange}{rgb}{1,0.5,0}
\newcommand{\todo}[1]{\textsf{\textbf{\textcolor{Orange}{[#1]}}}}

\newcommand{\PZ}[1]{\textcolor{black}{#1}}

\begin{document}
\title{RLOps: Development Life-cycle of Reinforcement Learning Aided Open RAN}

\author{Peizheng~Li,~\IEEEmembership{}
        Jonathan~Thomas,~\IEEEmembership{}
        Xiaoyang~Wang,~\IEEEmembership{}
        Ahmed~Khalil,~\IEEEmembership{} \\
        Abdelrahim~Ahmad,~\IEEEmembership{}
        Rui~Inacio,~\IEEEmembership{}
        Shipra~Kapoor,~\IEEEmembership{}
        Arjun~Parekh,~\IEEEmembership{}
        Angela~Doufexi,~\IEEEmembership{} \\
        Arman~Shojaeifard,~\IEEEmembership{}
        and~Robert~Piechocki ~\IEEEmembership{}
\thanks{Peizheng~Li, Jonathan~Thomas, Xiaoyang Wang, Ahmed~Khalil, Angela~Doufexi and Robert~Piechocki are with the Department of Electrical and Electronic Engineering, University of Bristol, UK (e-mail: \{peizheng.li, jonathan.david.thomas, xiaoyang.wang, oe18433, A.Doufexi,  R.J.Piechocki\}@bristol.ac.uk).

Abdelrahim~Ahmad and Rui~Inacio are with Vilicom UK Ltd. (email: \{Abdelrahim.Ahmad, Rui.Inacio\}@vilicom.com).

Shipra~Kapoor and Arjun~Parekh are with Applied Research, BT, UK (email: \{shipra.kapoor, arjun.parekh\}@bt.com).

Arman Shojaeifard is with InterDigital Communications, Inc. (email: arman.shojaeifard@interdigital.com).
}}



\maketitle 


\begin{tikzpicture}[remember picture, overlay]
  \node[minimum
  width=4in,fill=white!100,text=black,font=\normalsize, align=left] at ([yshift=-1cm, xshift=0cm]current page.north)  {This paper has been accepted for publication on IEEE Access (2022). DOI: 10.1109/ACCESS.2022.3217511};
\end{tikzpicture}
\begin{abstract}
Radio access network (RAN) technologies continue to evolve, with Open RAN gaining the most recent momentum. In the O-RAN specifications, the RAN intelligent controllers (RICs) are software-defined orchestration and automation functions for the intelligent management of RAN.
This article introduces principles for machine learning (ML), in particular, reinforcement learning (RL) applications in the O-RAN stack. Furthermore, we review the state-of-the-art research in wireless networks and cast it onto the RAN framework and the hierarchy of the O-RAN architecture. We provide a taxonomy for the challenges faced by ML/RL models throughout the development life-cycle: from the system specification to production deployment (data acquisition, model design, testing and management, etc.). To address the challenges, we integrate a set of existing MLOps principles with unique characteristics when RL agents are considered. This paper discusses a systematic model development, testing and validation life-cycle, termed: RLOps. We discuss fundamental parts of RLOps, which include: model specification, development, production environment serving, operations monitoring and safety/security. Based on these principles, we propose the best practices for RLOps to achieve an automated and reproducible model development process. At last, a holistic data analytics platform rooted in the O-RAN deployment is designed and implemented, aiming to embrace and fulfil the aforementioned principles and best practices of RLOps.
\end{abstract}


\section{Introduction}
\label{sec:introduction}
As the forefront of a mobile communication network, the Radio Access Network (RAN) directly interacts with the user equipment (UE). 
Its architecture has undergone profound changes in recent years, transitioning from monolithic to disaggregated architectures and from vendor-based to open-source solutions~\cite{bonati2020open}.
The disaggregation of the RAN is reflected in two vectors, one is the horizontal disaggregation of the network functions with open interfaces, and the other is the virtualization of hardware and software in vertical. To achieve efficiency dividends, allow for increased innovation, and also performance gain, O-RAN\footnote{In this
paper, O-RAN is taken as the reference architecture to demonstrate the validity of the proposed RLOps, but the principles can be applied to others RAN architectures.}
emerged from years of industry work in groups studying possible Open RAN trends (including the 3rd Generation Partnership Project (3GPP)). O-RAN is based on 3GPP new radio (NR) specifications, meaning it is 4G and 5G compliant, while the difference is merely the additional interfaces that are defined by O-RAN, focusing on a functional split called 7.2x.

The emphasis of O-RAN has been on \textit{openness} and \textit{intelligence} from the beginning \cite{alliance2020ran}, through which it intends to actively embrace the technological revolution brought by machine learning (ML). 
Within recent years significant research has been undertaken demonstrating the potential of ML within telecommunications, including channel estimation in massive multiple-input and multi-output (MIMO) systems \cite{soltani2019deep}, resource and service management in large-scale mobile ad-hoc networks \cite{forster2007machine} and mobile edge computation offloading and edge caching \cite{8493155}, to name but a few.
The introduction of this class of methods is supported through a number of avenues but perhaps most importantly through the definition of open interfaces and radio intelligence controllers (RICs).
Through their introduction, O-RAN provides the foundations by which ML models can be introduced into RAN. 
Thereby, facilitating the evolution of RANs from being static and stiff to data-driven, dynamically sensing and self-optimising. 
Notably, the exact mechanisms and procedures required to deploy cutting-edge ML solutions into production and realise their economic potential still need clarification.


As ML models and systems continue to mature they are experiencing increased adoption within a range of industrial settings.
The process through which they are developed and deployed is being formalised under the banner of MLOps~\cite{renggli2021data}.
Whereby, this is comparable to DevOps \cite{ebert2016devops} and emphasises similar best practices whilst considering the unique challenges which the relevance of data in ML model development introduces. 
In order to reliably and consistently bring the potential of ML to O-RAN, an operational platform implementing MLOps principles whilst considering the unique challenges of RANs is required.
These challenges pertain to its prominence as the critical national infrastructure and the highly dynamic nature of the platform. 
We place particular emphasis on the challenges of Reinforcement Learning (RL) within O-RAN due to the numerous applications that exist for it and its relative immaturity in terms of industrial applications.
Where we discuss key elements throughout the applications lifecycle including design considerations, challenges pertaining to training within the simulation and effectively monitoring live deployments, to name but a few. 
This pipeline and associated set of principles is coined \textit{RLOps}.

In this paper, we introduce the principles and best practices of RLOps in the context of an O-RAN deployment \cite{alliance2020ran}. 
To the best of our knowledge, this is the first work to systematically discuss the life-cycle development pipeline of ML, especially RL models in O-RAN and put forward a network analytics platform in accordance with RLOps. We explain the fundamental principles and highlight critical factors involved in RLOps. 
In Section \ref{sec:background}, we briefly introduce ML and RL, and the evolution of RAN and O-RAN architectures are given in detail. Next, we discuss some related applications of ML/RL in intelligent O-RAN, and correspondingly a series of challenges encountered in the development and deployment stages of ML/RL models. 
In Section \ref{sec: principles RLOps}, we elaborate on the principles of RLOps from the perspective of design, deployment and operations. We highlight the safety and security concerns in RLOps. In Section \ref{sec:best practice}, we put forward the effective routines and best practices of operating the aforementioned principles from the view of digital twins, automation and reproducibility.
Section \ref{sec:data engineering} illustrates the O-RAN related data analytics platform designed for achieving the principles and best practices of RLOps. 
Finally, Section \ref{sec:Conclusion} concludes this paper. Table~ \ref{table:LIST OF ACRONYMS} gives the list of used acronyms.
\section{Relevant Background}
\label{sec:background}

\subsection{Machine Learning in general}
\label{sec:ML in general}
Machine Learning (ML) is a branch of Artificial Intelligence (AI) concerned with learning from data, e.g. supervised learning (SL) and unsupervised learning (UL), or interaction, e.g. RL \cite{dunjko2018machine}.
In general, ML considers the utilization of an adaptive model parameterized by $\mathcal{\theta}$ with the intention of minimizing some objective function $\mathcal{J(\theta)}$. 
The exact objective function and form of the adaptive model depend on the exact formulation of the task (or set of tasks) we are interested in. 

Within SL tasks, we are typically presented with a dataset $\mathcal{D} = \{x_{i}: y_{i}\}_{\{i \in \mathcal{|D|}\}}$ comprising of feature vectors $x_i$, which are labelled $y_i$ which may refer to a discrete categories (cat or dog, for example) in a classification task or a real number if it is a regression task. 
Within this class of problem the objective is to learn a mapping $\mathcal{F}_{\theta}: \mathcal{X} \rightarrow \mathcal{Y}$.
A typical formulation of our objective function is the minimizing of Negative log-likelihood in the case of classification or the Mean Squared Error in the case of regression tasks. 

Like SL, in UL we are typically presented with a dataset $\mathcal{D} = \{x_{i}\}_{\{i \in \mathcal{|D|}\}}$, but in this case there are no labels. 
When presented with a task of this nature, we may be interested in clustering \cite{caron2018deep}, density estimation \cite{silverman2018density} or in dimensionality reduction for visualization \cite{Bishop2006}. 

RLs interaction with data is fundamentally different from other forms of ML. 
Typically, the problem is formalised as a Markov Decision Process (MDP), where this is defined by the tuple $<\mathcal{S, A, P, R, \gamma}>$.
Where $\mathcal{S}$ is the set of environment states, $\mathcal{A}$ is the set of actions that an agent performs, $\mathcal{P}$ represents the transition probability from any state $s\in \mathcal{S}$ to any state $s^\prime \in \mathcal{S}$ for any given action $a \in \mathcal{A}$.
$\mathcal{R}$ is the reward function that indicates the immediate reward received from the transition from $s$ to $s^\prime$, and $\mathcal{\gamma}$ is the discount factor that trades off the instantaneous and future rewards. The intention is to find a policy $\pi: \mathcal{S} \rightarrow \mathcal{A}$ which maximises the expected cumulative discounted reward $\mathcal{G}$~\cite{zhang2021multiagent} as defined in Equation \ref{defination of G}. 
\begin{equation}
    \mathcal{G} = \mathbb{E}[\sum\limits_{t\ge 0}{{{\gamma }^{t}}\mathcal{R}({{s}_{t}},{{a}_{t}},{{s}_{t+1}}|{{a}_{t}}\sim \pi (\cdot |{{s}_{t}}),{{s}_{0}})}]
    \label{defination of G}
\end{equation}
The process of finding this $\pi$ requires exploratory behaviours such that the agent can evaluate policies and learn about the MDP.
The parameterization of the adaptive model may vary; for example, in model-free algorithms, we may parameterise our $\pi$ directly or the state-action value $\mathcal{Q}$, or in model-based algorithms, we may learn a model of the MDP directly \footnote{Can be combined with model-free approaches.}.
\begin{table*}[htb]
\centering
\caption{LIST OF ACRONYMS}
\label{table:LIST OF ACRONYMS}
\begin{tabular}{llll}
\toprule
\textbf{Acronym} & \textbf{Definition} & \textbf{Acronym} & \textbf{Definition} \\ \midrule
3GPP     & 3rd Generation Partnership Project  & ML           & Machine Learning\\
5G       & 5th Generation Mobile Network       & mMTC         & Massive Machine-type Communications\\
AI       & Artificial Intelligence             & NE           & Network Elements\\
API      & Application Programming Interface   & Near-RT RIC  & Near-real Time RAN Intelligent Controller\\
BBU      & Baseband Unit       & NFV          & Network Function Virtualization\\
BS       & Base Station        & NLP          & Natural Language Processing\\
CAPEX    & Capital Expenditure                 & Non-RT RIC   & Non-real Time RAN Intelligent Controller\\
CPO      & Constrained  Policy  Optimization   & OFDM         & Orthogonal Frequency-division Multiplexing\\
CI/CD    & Continuous Integration (CI) and Continuous Deployment & OPEX & Operating Expense\\
CMDP     & Constrained MDP                     & O-RAN        & Open RAN\\

C-RAN    & Centralized (or Cloud) Radio Access Network & PL   & Processing Layer\\   
CSI      & Channel State Information  & PPO & Proximal Policy Optimization \\
CU       & Centralized Unit  & PRB   & Physical Resource Block\\

CV       & Computer Vision           & QoE          & Quality of Experience\\
DCA      & Data Collection Agents                    & QoS          & Quality of Service\\
DT       & Digital Twins                    & RAN          & Radio Access Network\\
DL       & Deep Learning              & RF           & Radio Frequency \\
DML      & Data Mediation Layer                       & RIC          & RAN Intelligent Controller\\
D-RAN    & Distributed RAN                       & RL           & Reinforcement Learning\\
DQN    & Deep Q Learning                & RNN          & Recurrent Neural Network\\
DRL      & Deep Reinforcement Learning                    & RRHs          & Radio Remote Heads\\

DSL      & Data Storage Layer       & RU           & Radio Unit\\

DU       & Distributed Unit         & SARSA          & State–action–reward–state–action\\

eMBB    & Enhanced  Mobile  Broadband    & SDN          & Software-defined Networking\\
ETSI    & European Telecommunications Standardization Institute                    & SL           & Supervised Learning\\

FCAPS   & Fault, Configuration, Accounting, Performance, Security         & SLA          & Service Level Agreement\\
FDD     & Frequency Division Duplexing & SMO & Service Management and Orchestration\\

IID     & Independent and Identically Distributed & SON & Self-Organizing Network \\
IP      & Internet Protocol        & SOTA         &  State-of-the-art  \\
IRS     & Intelligent Reflecting Surfaces  & TTI          & Transmission Time Interval \\
LSTM    & Long Short Term Memory               & UAV          & Unmanned Aerial Vehicle\\
LDPC    & Low-density Parity-check Code      & UE           & User Equipment\\  
MCTS & Monte Carlo Tree Search             & UL           & Unsupervised Learning\\ 
MDP     & Markov Decision Process    & URLLC        & Ultra-reliable low-latency Communications\\ 

MEC     & Mobile Edge Computing             & V2X          & Vehicle-to-everything\\
MIMO    & Multiple-input and Multiple-output               & VNFs         & Virtual Network Functions\\ 
&  & vRAN         & Virtual Radio Access Network\\
\bottomrule
\end{tabular}
\end{table*}

\subsection{Evolution of the RAN and O-RAN architecture}
\label{subsec:oran architecture}
A typical mobile communication network mainly comprises a RAN, a transport network and a core network. The RAN gives the UE access to the core network, this subsequently provides the services to the user. The transport network implements the IP routing and IPSec functionality that securely connect the different network elements and network domains of the mobile network, thus allowing for full end-to-end functionality.

From 1G to 5G, the evolutionary trends of communication systems are the modularity and virtualization of decoupled network functionalities. For instance, the core network embraces x86 platform universal servers and performs network function virtualization (NFV), where the slicing of the core network embodies this feature. However, due to the complexity of antennas, the Remote Radio Heads (RRHs), and the Baseband Units (BBUs) in RAN, the functionality decoupling of RAN is slower than the decoupling of the transport network and core network. Three distinct structural improvements have been proposed in the evolution of RAN, namely the distributed RAN (D-RAN), the centralized (or cloud) RAN (C-RAN), and vRAN.
The RRHs and BBUs are co-located in D-RAN at every distributed cell site. The RRHs and BBUs communication are provided by the proprietary interfaces. Cells are connected back to the core network through the backhaul interface.
In C-RAN, all BBUs are further concentrated into the centralized BBU pool for cloudification, and every site merely keeps antennas and RRH. RRHs and the centralized BBU are connected with fronthaul. Centralized BBUs bring the convenience of cell deployment and maintenance and significantly reduce the CAPEX and OPEX. vRAN decouples the software and hardware by NFV, where the BBU is virtualized on x86 servers \cite{faisal_2021}. 
In 3GPP 5G NR related specifications, the above Base Station (BS) components are reorganised into the centralized unit (CU), distributed unit (DU) and radio unit (RU) entities, with their deployment following a flexible topology. CU and DU play the role of BBU, and the RU functions the converting between the signals and radio frequency (RF). A more comprehensive review regarding the details of interfaces and radio evolution is presented in \cite{wypior2022open}. In the meantime, all the hardware design, specialized software development and intellectual properties of the RAN-related components are still proprietary.

Network operators expect to obtain decoupled, standardized RAN hardware and open-source operating software to relieve current vendor restrictions. Consequently, the O-RAN alliance was founded in February 2018.
Its ambitious mission is to reshape the RAN industry, building future RANs on a foundation of virtualized network elements, white-box hardware, and standardized interfaces. The core principles of O-RAN are intelligence and openness, which will lead the direction beyond 5G and 6G.
 \begin{figure*}[t]
    \centering
    \includegraphics[width=0.85\textwidth]{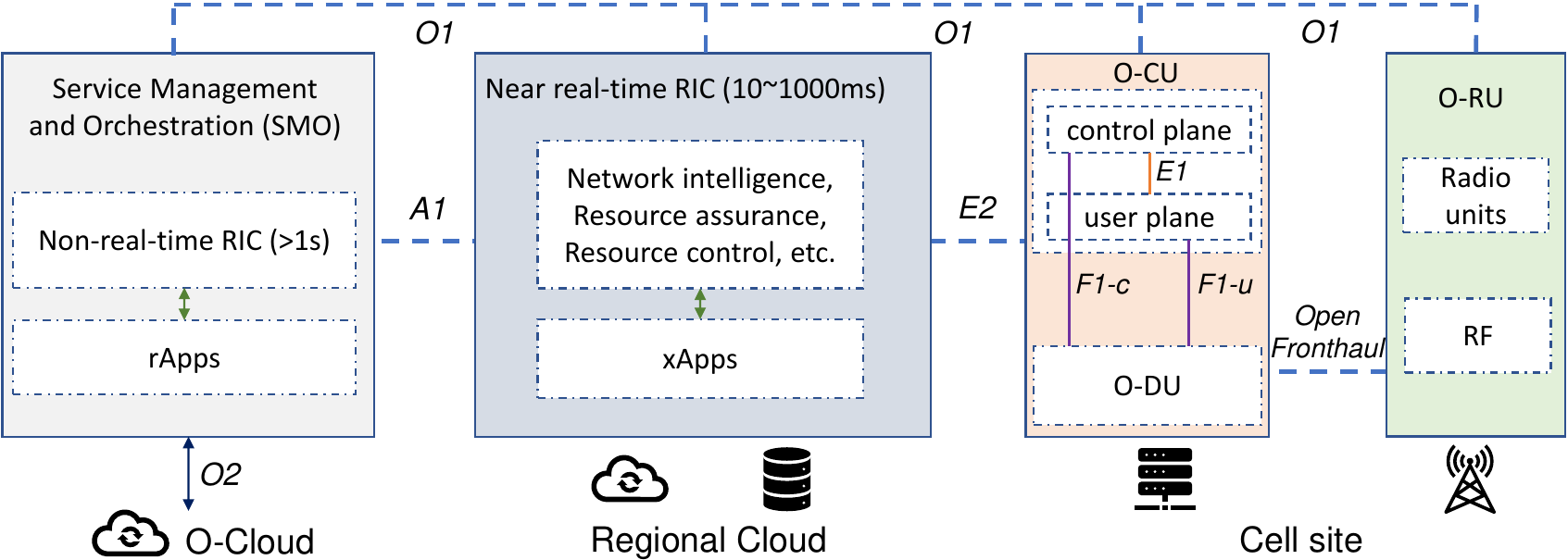}
    \caption{One example architecture of O-RAN.}
    \label{fig:ORAN_architecture}
\end{figure*}

Fig.~\ref{fig:ORAN_architecture} demonstrates one example architecture of O-RAN. 
O-RAN architecture follows 3GPP architecture and interface specifications, while its NFV is as consistent as possible with European Telecommunications Standards Institute (ETSI).
The service management and orchestration (SMO) function in O-RAN has been designed to provide network management functionalities for the RAN and may also be extended to perform core management, transport management, and end-to-end slice management. Meanwhile, the SMO connects with O-Cloud through the O2 interface. O-Cloud is a cloud computing platform comprising a collection of physical infrastructure nodes that meet O-RAN requirements to host the relevant O-RAN functions, the supporting software components and the appropriate management and orchestration functions \cite{alliance2020ran}.
One important functionality provided by SMO is the Non-RT RIC designed to implement automated policy-based optimization activities by running ML models. The Non-RT RIC links towards Near-RT RIC via A1. The Near-RT RIC controls and optimizes the functions of CU and DU through the E2 interface. Meanwhile, third-party, microservice architecture-based applications can also be loaded into the Non-RT RIC and Near-RT RIC through rApps and xApps, respectively, to perform data-driven optimization behaviours. 
In this process, E2 can be leveraged to access the radio node data, and these data can be fed into RICs for ML model training. The CU connects to or controls one or more DUs via the F1 interface. Similarly, one DU connects to at least one RU through the open fronthaul plane. 
The CU/DU stack hierarchically handles operations of different timescales, while the RU manages and controls the most fundamental RF components and the physical layer in every RU deployment site. All functions of the O-RAN, including the Near-RT RIC, CU, DU and RU, are connected to the SMO through the O1 interface for FCAPS support.

It is noticeable that three control loops involving system parameters and resource allocations are defined in O-RAN. ML solutions can be adopted in any loop based on the time-sensitivity of tasks, in which loop1 handles operators at the time scale of TTI level (<10 ms) for those scenarios that emphasize real-time like the radio resource control and allocation happened in between DU and RU; loop 2 operates in the Near-RT RIC which deals with tasks operating within 10-500 ms. It mainly aims to the O-RAN internal resource control, which RICs perform; loop 3 operates in the Non-RT RIC to process tasks greater than 500 ms.

\subsection{ML/RL applications in O-RAN}
\label{sec: ML/RL applications}
ML is undoubtedly the most remarkable technological progress in recent years. From CV \cite{akhtar2018threat}, NLP \cite{otter2020survey} to robotics \cite{wu2020reinforcement}, gaming \cite{silver2017mastering}, e-commercial \cite{yang2019aligraph} and biology \cite{jumper2021highly} etc. ML applications in almost every technical field have made marvellous achievements. Also, the upcoming O-RAN through the introduction of intelligent programmable RIC enables the RAN to have a mechanism to use emerging learning-based technologies to automate network functions, improve network efficiency, and reduce operating costs.
In O-RAN, the initially closed internal radio resources are opened and controlled by unified RICs. 
That brought some profound changes to communications studies.
\begin{enumerate}
    \item With higher mobile edge computing (MEC) capability, O-RAN enables interaction with end-users, such as directly perceiving end-users behaviours and responding to them so that the optimization of the network can be completed from a more fine-grained and more direct user model analysis way, without the need to perform it in the core network or the centralized cloud. 
    \item O-RAN can significantly help the further promotion of 5G. As often mentioned, the main goals of 5G are enhanced mobile broadband (eMBB), ultra-reliable low-latency communications (URLLC) and massive machine-type communications (mMTC) \cite{lien2011toward}. Due to the complex and diverse environmental conditions faced by 5G networks, it is necessary to allocate resource blocks with network slicing to meet task requirements for different application scenarios. The introduction of O-RAN makes a dynamic, learning-based slicing mechanism possible. Therefore, deep learning-based adaptive slicing, the collaboration of SDN and NFV, is becoming a research hotspot.
    \item RICs provide a platform for third-party applications deployment, including ML models, enabling the rapid development and deployment of innovative ideas and algorithms.
\end{enumerate}

\begin{table*}[htb]
\centering
\caption{\PZ{A survey of the SOTA works regarding DRL applications under the O-RAN context. According to the attributes of these tasks, we divide them into four categories: network slicing, scheduling, and splitting; connection management; resource allocation and xApps related.}}
\label{table:RL and O-RAN}
\begin{tabular}{p{2.8cm} p{4.5cm} p{2.5cm} p{5.5cm}}
\toprule
\textbf{Paper} & \textbf{Task} & \textbf{Algorithm} & \textbf{Gains} \\ \midrule
Abedin et.al. \cite{abedin2022elastic}    & Elastic O-RAN slicing   & Actor-Critic  & 50\% at severed devices  \\

Bonati et.al. \cite{bonati2020intelligence}    & RAN slicing allocation and scheduling   & PPO           & 20\% at spectral efficiency \\

Filali et.al. \cite{filali2022communication}    & RAN resource slicing   & Double DQN & Robust and efficient performance for URLLC services\\

Pamuklu et.al. \cite{9500721}    & Dynamic function splitting   & SARSA, Q-learning   & Efficiency on renewable energy usage and cost\\

Polese et.al. \cite{9814869}    & RAN slicing allocation and scheduling  & PPO           & \makecell[l]{Improved PRB ration and throughput;\\ smallest buffer occupancy for the MTC traffic; \\
30\% at fewer data requirements of online-training}\\ \midrule

Lien et.al. \cite{lien2021session}    & Session management for URLLC   & \makecell[l]{SARSA, Q-learning\\ and double Q-learning}   & Enabling the gNB to grant a new URLLC session or not\\

Mollahasani et.al. \cite{mollahasani2022energy}    & Dynamic DU selection   & Soft Actor-Critic  & 50\% at energy efficiency \\

Orhan et.al. \cite{9679991}    & Optimisation of  user-cell association   & Graph RL & 10\% - 140\% at throughput, cell coverage or load balancing\\

Wang et.al. \cite{wang2021self}    & CU–DU resource assignment   & Neural MCTS   & 5.70\% - 12.95\% at resource utilisation efficiency\\

\midrule

Iturria-Rivera et.al. \cite{iturria2022multi}    & Power and radio resource allocation   & Multi-agent DRL  & Higher energy utilization and throughput \\

Mungari \cite{mungari2021rl}    & Radio resource management  & -           & Dynamic resource allocation based on traffic flow \\

Zhang et.al. \cite{zhang2022team}    & Power and radio resource allocation   & Team DQN & Higher system throughput and lower packet drop rate\\
Giannopoulos et.al.\cite{giannopoulos2022supporting} & Power allocation & Multi-agent DRL & 72.3\% at energy efficiency\\
\midrule

Bonati et.al. \cite{bonati2022openran}    & Facilitate the xApps development   & -   & \makecell[l]{An integrated platform for data collection, model \\development and xApps building}\\

Zhang et.al. \cite{zhang2022federated}    & Multiple xAPPs coordination   & Federated RL  
& 11\% higher throughput and 33\% lower delay \\ 
\bottomrule
\end{tabular}
\end{table*}

The O-RAN use case whitepaper \cite{alliance2020ran} described some of the AI-based deployment targets, such as service level agreement (SLA) assured 5G RAN slice, context-based dynamic handover management for vehicle-to-everything (V2X), traffic steering, and flight path based dynamic unmanned aerial vehicle (UAV) resource allocation etc., while we believe the potential of AI-enabled O-RAN is far more than that. The state-of-the-art communication system embodies a feature of hierarchical and self-contained functions. All functions are interconnected with standardized interfaces. For instance, the signal undergoes a series of units from the transmitter to the receiver, such as modulation, coding, demodulation, de-noising, and corresponding channel measurement.
Each unit has a well-defined mathematical model that can approach the Shannon limit, and it can be considered that a single unit has achieved its local optimum.
However, there are significant challenges in the analysis and optimization of cross-units. If the whole of the above units is regarded as the optimization object, then this kind of global or multi-objective optimization is currently challenging to achieve \cite{ouyang2021next}.
The combination with ML/RL various learning paradigms makes O-RAN have the potential for this overall or multi-objective optimization revealed in some advanced research. For instance, in the physical layer, the DL-based OFDM receiver can achieve accurate channel estimation using fewer pilot signals \cite{van2019deep}; the end-to-end learning of communication systems has been realized in an autoencoder way which shows advantages in synchronization, equalization and dealing with hardware impairments such as non-linearities \cite{8445920}; the BS downlink channel state information (CSI) in frequency division duplexing (FDD) massive MIMO system can be inferred by DL with feeding the downlink CSI under certain conditions \cite{yang2019deep}; under the premise of imperfect CSI, the design of hybrid massive mimo digital precoder and analog combiner based on RL \cite{wang2020precodernet}; and a variety of DL-based LDPC decoding solutions under harsh noise \cite{wang2018unified}.
In the network layer, the learning-based algorithms shape the SON with dynamic resource allocation properties like automated networking, slicing, dynamic spectrum sensing, random access channel, \PZ{5G cooperative communication and resource allocation \cite{guo2022cooperative} \cite{bashir2019optimal}}, and load balancing optimization in the network layer. 
It is to be noted that with the O-RAN stepping into the market gradually, some DRL-based optimization cases targeting O-RAN's features are beginning to appear.
\PZ{A survey of  state-of-the-art (SOTA) works regarding DRL applications embracing the O-RAN is shown in Table~\ref{table:RL and O-RAN}. According to the attributes of these tasks, we divide them into four categories: network slicing, scheduling, and splitting; connection management; resource allocation and xApps development related. The corresponding algorithms and gains are also detailed in this table.}


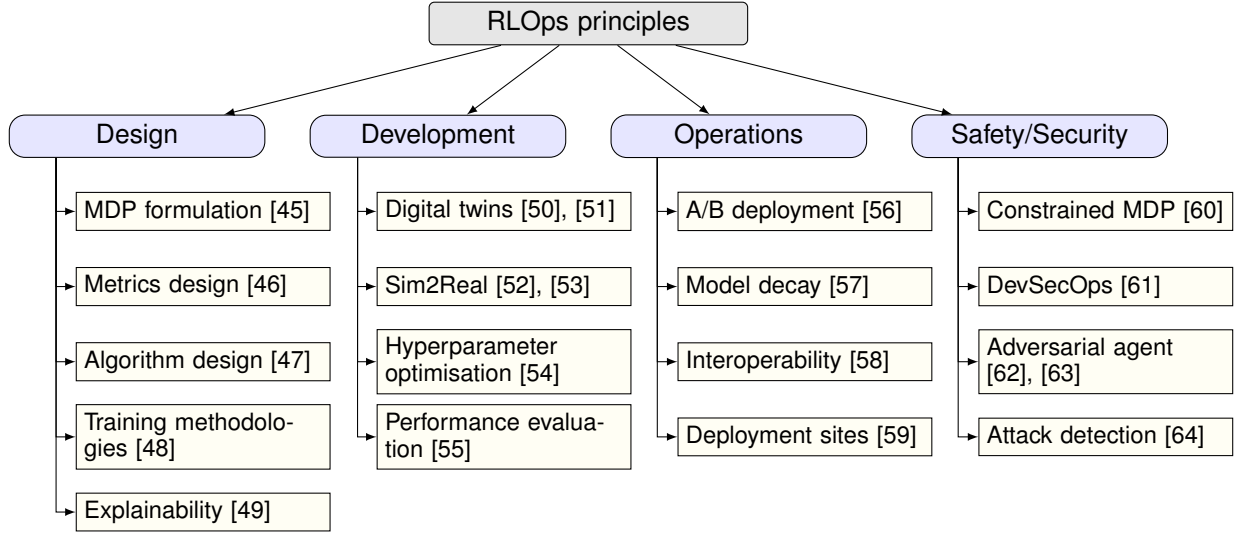
\begin{figure*}[t]
\centering
\begin{tikzpicture}[
  level 1/.style={sibling distance=40mm},
  edge from parent/.style={->,draw},
  >=latex]
 
\node[root] {RLOps principles}
  child {node[level 2] (c1) {Design}}
  child {node[level 2] (c2) {Development}}
  child {node[level 2] (c3) {Operations}}
  child {node[level 2] (c4) {Safety/Security}};
  
[font=\small]
\begin{scope}[every node/.style={level 3}]
\node [below of = c1, xshift=25pt] (c11) {MDP formulation \cite{padakandla2021survey}};
\node [below of = c11] (c12) {Metrics design \cite{jordan2020evaluating}};
\node [below of = c12] (c13) {Algorithm design \cite{arulkumaran2017deep}};
\node [below of = c13] (c14) {Training methodologies \cite{moerland2020model}};
\node [below of = c14] (c15) {Explainability \cite{puiutta2020explainable}};

\node [below of = c2, xshift=25pt] (c21) {Digital twins \cite{fuller2020digital,9374645}};
\node [below of = c21] (c22) {Sim2Real \cite{zhao2020sim, 9158349}};
\node [below of = c22] (c23) {Hyperparameter optimisation \cite{springenberg2016bayesian}};
\node [below of = c23] (c24) {Performance evaluation \cite{li2017deep}};

\node [below of = c3, xshift=25pt] (c31) {A/B deployment \cite{gauci2018horizon}};
\node [below of = c31] (c32) {Model decay \cite{DBLP:journals/corr/abs-2004-05785}};
\node [below of = c32] (c33) {Interoperability \cite{medel2016modelling}};
\node [below of = c33] (c34) {Deployment sites \cite{alwarafy2021deep}};

\node [below of = c4, xshift=25pt] (c41) {Constrained MDP \cite{altman1999constrained}};
\node [below of = c41] (c42) {DevSecOps \cite{myrbakken2017devsecops}};
\node [below of = c42] (c43) {Adversarial agent \cite{chen2019adversarial, elderman2017adversarial}};
\node [below of = c43] (c44) {Attack detection \cite{pattanaik2017robust}};

\end{scope}

\foreach \value in {1,...,5}
  \draw[->] (c1.195) |- (c1\value.west);

\foreach \value in {1,...,4}
  \draw[->] (c2.195) |- (c2\value.west);

\foreach \value in {1,...,4}
  \draw[->] (c3.195) |- (c3\value.west);
  
\foreach \value in {1,...,4}
  \draw[->] (c4.195) |- (c4\value.west); 
\end{tikzpicture}
\caption{This figure lists the critical elements involved in the principle of RLOps. That is the design, development, operations and safety/security, and then we further break down each element into the high-level taxonomy of considerations and methodologies.}
\label{fig:RLOps sorting}
\end{figure*}
\subsection{Challenges of ML/RL developing in O-RAN}
\label{subsec:challenges of ML/RL in ORAN}
Although the hierarchical structure and decoupling characteristics of O-RAN have brought the benefits of supplier diversification, this also brings in higher complexity in O-RAN deployment. On the other hand, developing and deploying a suitable intelligent model in O-RAN may lead to practical engineering technology problems. 
Considering the most vigorous ML domains in CV and NLP, some standard data sets are generally used to evaluate the performance of the developed ML algorithms. These algorithms are designed to target the features of the given training sample. 
For example, for the image sample, the initial features of the image space are extracted from adjacent pixels through various convolution operations, and then through various sophisticated network structures such as AlexNet, VGG, and ResNet, the features are further refined. The mapping from the training space to the target space is accurately constructed. 
In the booming RL field, whether it is gaming or robot control, the development of related algorithms is basically carried out under a standard toolkit like OpenAI gym \cite{brockman2016openai}. A noteworthy phenomenon is that the fields mentioned above benefit from the support of solid mathematical models and complete underlying software. The development of involved ML has been systematically transformed into a near-standard industry.
These models corresponding to different application scenarios are well defined, making the goal of algorithm development precise and the whole process controllable.

Turn our attention to the application of ML in O-RAN. The state-of-the-art progress made by the current O-RAN alliance is summarised as follows.
(1) The programmable and expandable RIC modules are introduced into the O-RAN architecture, and an interface for data collection within the network is defined.
(2) With the clarification of the structure definition, a series of ambitious optimization or control goals for resource, traffic flow, and power consumption have been proposed.
(3) The workflow of using SL and RL has been standardized.
However, the above progress only reflects the possibility of O-RAN embedded ML in a broad and macro sense. Specific to the realistic implementation of the ML models, we will encounter a rather complicated situation. We further consider issues of ML in O-RAN from algorithms development and deployment angels, respectively.
From the view of algorithms development, we summarised the potential issues below:
\begin{enumerate}
\item In O-RAN, data related to model training is difficult to obtain and process. Even the standard interfaces defined in the O-RAN architecture, such as E2, can access DU, CU and other components to collect information inside the network. This data comes, by default, in raw format and without a schema that is not suitable to be directly consumed by ML/RL algorithms. If we intend to use this field information to train the model, the cost of data collection will be very high.

\item For different optimization goals, the required data for neural network training is heterogeneous. The attributes or patterns of various types of data hiding are elusive.
For example, for radio traffic, the data flow as a whole is usually non-Euclidean. In some RU-distributed sites, the data does not meet the characteristics of independent and identically distributed (IID), and some data sets have very strong temporal correlations, while the correlation of other data sets is more reflected in the spatial domain. That will pose challenges to the subsequent data processing methods and feature extraction schemes, affecting the overall neural network structure design.

\item Some global optimization problems demonstrate the applicability of RL. That poses other challenges for establishing the connection between O-RAN and RL. These challenges are often not about the RL algorithm itself but how to abstract the problem to be solved into the RL framework and define the RL-related environment, action, state, and reward. For instance, the training issue comes along with high-dimensional state and action spaces; the availability of offline models trained from historical logs; the feasibility of online model training but with limited samples or partial observations; the large reward delay or vanishing in RANs; the complexity of multi-agent RL scheme for optimization problems across multiple RANs \cite{dulac2019challenges}. 

\item The RAN is the entrance to the entire wireless network and is closest to UEs. Therefore, the data flow in O-RAN is inevitably directly related to UEs. If we want to use these data streams to train neural network models, new requirements will be put forward for the privacy protection of the UEs and the desensitization of related data.

\item O-RAN supports multi-vendor third-party ML/RL applications and increases the complexity of the processes and activities related to the network management plane, which may result in action conflicts in execution, especially when resource allocation is involved.  The action coordination ought to be considered in the process of model training \cite{zhang2022team}. 
\end{enumerate}



We have introduced that xApps are connected to Near-RT RIC in O-RAN as the host of trained ML models. These trained models are pre-stored in O-cloud and managed by the SMO. However, from the view of model deployment, the above system is not enough to overcome the problems that may arise after the model is deployed in the field. On the one hand, the models obtained by SL were trained by specific data sets. After deploying these mature models, one possible consequence is that the sample data characteristics in the model deployment area are inconsistent with the characteristics of the original training set, which will result in model failure; that is, the expected results cannot be correctly received, as the model can not respond to the input features. On the other hand, as time changes, the external environment changes continuously for the RL model, which will make the initially trained policy no longer suitable.
That puts forward new requirements for model management, update and maintenance, and we must look at O-RAN and its ML models from a more holistic perspective.

\section{Principles of RLOps}
\label{sec: principles RLOps}
\subsection{Brief introduction of  MLOps}
MLOps is defined as a set of practices that combines ML, DevOps and data engineering, aiming to deploy and maintain ML models in production reliably and efficiently. It can be seen as delivering ML applications through DevOps, with additional attention to data and models. MLOps performs the idea of \textit{automation} and \textit{acceleration}. \textit{Automation} means automating the ML pipeline from data to model for continuous training, as well as automated CI/CD for ML applications. \textit{Acceleration} means to increase the speed of delivery while maintaining the quality of service for ML applications~\cite{43146}.

An MLOps pipeline usually consists of the following elements:
\begin{enumerate}
    \item Data preparation and model design.
    \item Model testing and validation.
    \item Model integration, delivery and monitoring.
    \item Continuous training and CI/CD.
\end{enumerate}

Similar to DevOps, MLOps is an iterative approach. The change in developing requirements, the evolution of the deployment environment, and the alerts raised by monitoring the deployed model would trigger the execution of the pipeline to guarantee the quality of ML applications.

\subsection{Motivation for RLOps}
 MLOps is the general principles and practices of continuous delivery and automation pipelines in ML. Considering the increasing applications of RL in communication networks, we study the “RLOps” principles to deliver the value of RL to the industry.

 \begin{figure}[t]
    \centering
    \includegraphics[width=0.33\textwidth]{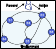}
    \caption{An example of RL.}
    \label{fig:MDP}
\end{figure}

RL differs from other ML approaches in several ways, which brings the need for more targeted principal sets. As shown in Fig.~\ref{fig:MDP}, \textit{Data \& Environment}, \textit{Agent} and \textit{Reward} are the key distinctions considered in the design and delivery of RL applications.
\begin{itemize}
    \item \textit{Data \& Environment}. Data is considered the backbone of ML practices. In RL, data is from agents interacting with environments (online RL) or pre-collected datasets (offline RL)~\cite{levine2020offlineRL}. For online RL, the interaction and learning from live environments (in our case, live communication networks) brings additional risks (as we discussed in Section \ref{subsec:challenges of ML/RL in ORAN}), which is infeasible in some cases. Hence, the idea of digital twins (DT) has been brought up as a promising solution to the environment and data issue of RL practices~\cite{9374645}, providing a controllable, reliable and easily accessible simulation environment. We elaborate it in Section \ref{sec:dt}. Furthermore, considering other real-data hungry cases, communication networks bring challenges to environment access, training data acquisition and model validation. A network analytics platform with automated data collection, pre-processing, model validation and management abilities are proposed and discussed in Section \ref{sec:data engineering}. 
   \item \textit{Agent}. Agents are the core of RL problems, interacting with the environment and following their policies. The policy that agents perform is the brain of MDP solutions, as counterparts for “models” in SL and UL. The general principles for developing and deploying ML models also apply to RL models, including model analysis, testing and monitoring. 
    \item \textit{Reward}. The reward is unique to MDPs. It represents the goal of RL, which is essential information to have in model design and deployment. Unlike “labels” as intrinsic features of data in SL, rewards reflect the expected behaviour of agents. In RL applications, reward design is always part of the problem formulation, which requires special attention in RLOps.
\end{itemize}

\PZ{As illustrated in Table~\ref{table:RL and O-RAN}, 
a large amount of work has been demonstrated for the specific approaches of developing RL-based model in O-RAN recently. However, we believe that on top of these use cases, some common considerations and issues need to be solved, at least, to be realised. By doing that, we expect some general principles of developing and deploying RL models can be summarised to realise true life-cycle management and continuous integration and delivery of such models. Hopefully, a more realistic and affordable RL developing pipeline can be put forward to fulfil the above objective rather than developing case by case. That is the essential intention of this paper.}

To effectively deploy RL applications requires careful navigation through a wide range of decisions, from problem formulation to algorithmic choices to the selection of monitoring metrics, to name a few. 
In an attempt to demystify these decisions, we introduce a non-exhaustive list of ''RLOps'' principles and observations, which we consider helpful in realizing the potential RL promises.
We hope to provide distinct but complementary ideas for RLOps to what may be expected in MLOps and DevOps.
For an overview of key considerations and principles for MLOps please refer to \cite{8258038}.

\begin{figure}[t]
    \centering
    \includegraphics[width=0.45\textwidth]{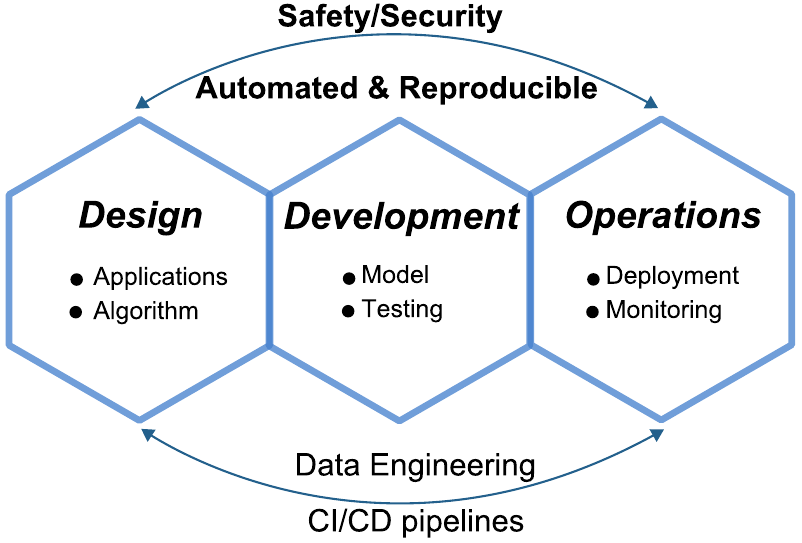}
    \caption{RLOps diagram. This diagram demonstrates the development cycle of applications.}
    \label{fig:RL_Ops}
\end{figure}

We introduce principles of RLOps under the application development cycle introduced in Fig.~\ref{fig:RL_Ops}\footnote{Inspired and adapted from https://ml-ops.org/img/mlops-loop-en.jpg.}.
Below we talk about the three parts: design, development and operation. We will also elaborate on the safety and security concerns related to these three parts. A summary of the high-level taxonomy of considerations and methodologies involved in the RLOps principles is shown in Fig.~\ref{fig:RLOps sorting}.

\subsection{Design in RLOps}
\textit{$*$ The challenges of design in RLOps lie in the appropriate task formulation and algorithm selection specific to the dynamic environments.}
\subsubsection{Task Formulation}
Consider the arrival of a new task that takes the form of sequential decision-making, as such RL is likely to be a good solution. 
Examples of these tasks are given in Section~\ref{sec:introduction} including handover and interference management, to name but a few \cite{9185035}. 


An integral step to build a solution based on RL is to formulate the given problem as an MDP.
The formulation of MDP affords many degrees of freedom. Each design decision should be considered carefully, as the form of the MDP will dictate a number of algorithmic decisions.
Basic elements to consider include the number of agents, the representation of actions the degree of stochasticity of the environment. 
For example, if the problem requires distributed decision-making, a stochastic game  \cite{shapley1953stochastic} may be an appropriate formulation; If the hierarchical representation of the action space is possible, and options framework \cite{florensa2017stochastic} may be possible.



As part of the task formulation phase, it is useful to consider evaluation metrics and baselines that are suitable for the task, where these baselines may be existing solutions.
This will smooth out the test, validation and monitoring phases in RLOps, and potentially provide a fail-safe if the RL application begins to behave erratically. 

\subsubsection{Algorithm}
\label{design:algorithm}
 
As discussed in the above section, decisions on the formation of MDP directly impact the form of the solution. Some of the design practices are listed below, but many other general rules exist. \cite{rao2020make} provides a good analysis of the impact of some design choices on specific RL algorithms.
\begin{itemize}
    \item If the action space is continuous, policy gradient-based approaches are likely to be a good option. Some discretisation could also be applicable, such as Q-Learning variants.
    \item If the state is non-markovian, recurrency can be introduced through stacking previous states\cite{mnih2013playing} or RNN structures like LSTM \cite{hochreiter1997long}.
    \item Particularly, small state-action spaces may be amenable to tabular approaches \cite{sutton2018reinforcement}, which provides a higher degree of interpretability over methods that use function approximation.
\end{itemize}

In the design phase, the training strategies and tricks are also worth considering to tackle problems like \textit{model generalisation} and \textit{training difficulty}.
Considering potential varying environments for deploying RL solutions, we are interested in how the model generalizes and as such our training methodologies should reflect this.
It has become a consensus in RL research that models trained within limited instantiation of environments do not generalize well~\cite{zhao2020sim}.
The utilization of methods like ``Domain Randomisation'' is essential for model generalisation. As for the training difficulty, the utilization of a training curriculum, where a series of increasingly complex tasks are presented to the agent with the intention of easing learning on difficult tasks \cite{narvekar2018learning}. Imitation Learning is another approach to ease the learning process, \cite{hussein2017imitation} where an agent is pre-trained with the pre-collected dataset containing expert behaviours \cite{levine2020offlineRL}. \PZ{In \cite{valencia2019using}, a binary neural network using neuroevolution is presented to simplify the inference model.}

In addition to these design choices, we may wish for our algorithm to possess other characteristics. For example, we may want its decisions to be explainable or for it to be aware of its uncertainty with regard to its state. 
That moves us to other sub-areas of RL researches like \textit{Explainability} \cite{puiutta2020explainable} and \PZ{\textit{Bayesian RL} \cite{ghavamzadeh2016bayesian} \cite{strens2000bayesian}} which deals with these concepts that are important from business, legislative or even safety perspectives.

\subsection{Development in RLOps}
\textit{$*$ The reliable platform and consistent data stream for model training and testing are critical challenges of development in RLOps.}

Once elements of the \textit{design} have reached sufficient maturity levels, steps can be taken towards formally developing the application's capabilities. 
This process involves creating the experimental environment (which will likely be based on a DT), model training, and performance optimization.

\subsubsection{Model}
The algorithmic approach defined in the Section \ref{design:algorithm} provides the general structure and algorithm for the model. 
The next step is to develop the necessary code for the agent. 
To code everything from “scratch” may seem reasonable but may lead to significant engineering expense for limited gain, especially when a wide array of readily available open-source libraries provide high-quality implementations of a range of SOTA algorithms exist\footnote{Ray and OpenAI Baselines, to name but a few.}. 

Training RL applications in a time-efficient and comprehensive manner requires accessibility to a high fidelity simulator - where a DT will likely be a good fit for this\footnote{The general structure of which should follow that mandates by OpenAI Gym for consistency with other open-source platforms.}.
If we take a pessimistic viewpoint, the DT or any simulator is an approximation to the real world, and as such, there will be inconsistencies in behaviour that may, in the worst case, lead to testing values being inconsequential as the differences are so profound that the policies are not transferable. 
This is a \textit{Sim2Real} challenge and is considered in more detail in Section \ref{sec:dt}. 

Once an effective algorithmic and simulation approach has been developed to address this challenge, the next major obstacle in the model development process is hyperparameter optimization, which is an arduous and time-consuming process.
In the interest of efficient allocation of resources, this process will benefit from automation.

\subsubsection{Testing}
In the life circle of DevOps, testing is essential to ensure the performance of software systems. Code sanity testing, unit testing and integration testing are commonly used to validate the software iteration. In MLOps \cite{google}, the scope of testing extends to data and models. Here we re-consider testing in the context of deep reinforcement learning (DRL) in future O-RAN.

Once a DRL model has been trained, we require functionality within our pipeline to evaluate the model’s capabilities.
For trained DRL models, testing should consider multiple model attributes to give a comprehensive evaluation of the models' performance. Some dimensions are also considered in MLOps, such as the model relevance and accuracy, the robustness to noise, the generalization ability, and ethical considerations \cite{renggli2021data}. Other challenges are unique to DRL models, for example, the ability of a DRL model to prioritize useful experiences during learning, to choose long-term beneficial actions, to respond to uncertainty, stochasticity, and environmental changes, and avoid unintended behaviour, etc. The testing and validation of DRL models regarding the dimensions mentioned above remain an open question, leaving space for future work. DT might play an important role in the testing procedure since it is an environment in which we have complete control. Manual testing might be required in some use cases. In addition, model interpretability and explainability are of great importance from the perspective of both developer and network service providers, which should be considered during testing. Considering possible network attacks and security challenges, an adversarial attack should also be integrated into the model testing workflow.

\subsection{Operations in RLOps}
\textit{$*$ The challenges of operations in RLOps lie in the agile and effective monitoring and identification of errors of RL models among different deployment sites.}

Assuming that a model has passed all required testing and validation steps and has been containerized according to system requirements, the obvious next consideration should be for model deployment and the associated systems required to support and maintain it.
This process will include consideration of the deployment location and monitoring with the intention of providing functionalities for continuous improvement. 

\subsubsection{Deployment}
A fundamental issue (which is discussed at more length in Section \ref{sec:monitoring}) is that of discrepancies that exist between development and production environments. 
This problem is likely to be ever-present and difficult to quantify.
As such, other safeguards are likely required to mitigate this risk before wide-scale deployment. 
An example of this could be using software development practices like alpha-beta type deployments to limit the potential impact on end-users whilst getting an empirical measure of application performance.


The environment in which the agents are deployed tends to be highly dynamic, where changes are likely to alter network behaviours. 
These changes may include internal factors like device configurations and the deployment of other applications or external factors like changes in user behaviour or seasonal phenomena that may affect wireless propagation characteristics. 
The manifestation of this dynamic environment is a modification to the underlying MDP, and the performance of RL agents will likely degrade accordingly.
This issue is one of the concept drifts \cite{DBLP:journals/corr/abs-2004-05785}. 
The implication of the dynamic nature of the deployment environment is that the performance of deployed models may reduce over time. 
This general phenomenon is known as \textit{Model Decay} and will be observable through the agent's reception of the reward. 
This impact can be mediated through periodic re-training if the reward drops below some pre-defined threshold.
An alternative approach is to enable online training, but this does come with risks, most notably the requirement for exploration. 
An additional risk that may arise is non-stationarity \cite{DBLP:journals/corr/abs-1906-04737}, which is a consequence of a deployment consisting of multiple RL agents constituting a Multi-Agent system. 
Non-stationarity arises when multiple agents are learning policies simultaneously, resulting in uncertainty regarding environment behaviours as state transitions are implicitly dependent on other agents.

To enable interoperability on differing base computational platforms all applications will need to be containerised with their associated internal dependencies for deployment with a platform like Kubernetes \cite{medel2016modelling}. 
A well-defined REST application programming interface (API) will allow for communication of information between entities such that applications can obtain external information that they require for operation and so that monitoring can be performed and decisions can be made pertaining to applications. Communication between disparate systems within the O-RAN architecture naturally raises considerations for model deployment location. 
By selecting an appropriate location (be that topological or cloud vs edge) and control loops described in Section \ref{subsec:oran architecture}, application performance and the wider performance of the network may be improved, where benefits are related to reduced inference time and a reduction in network traffic due to co-location of applications with their dependencies. 
These decisions may be particularly important for applications that require very low latency for effective operation.

\subsubsection{Monitoring}
\label{sec:monitoring}
Through a collection of Key Performance Indicators (KPIs), the efficacy of an RL agent can be monitored. 
This information enables decisions pertaining to the application to be made in an informed manner. For example, if an agent is underperforming, it may be desirable to re-train or even replace the agent with an alternative solution. 



Monitoring and evaluating RL application performance in the real world is critical to determining whether or not the application is providing benefit, but this is likely to be challenging. 
Simple measures like cumulative reward can be utilized but are susceptible to issues like reward hacking \cite{DBLP:journals/corr/AmodeiOSCSM16} and do not provide relative measures compared to other methods. 
The most thorough approach from a network operator's perspective may be to have human oversight of the decisions that agents are making, but this is not scalable and is likely to be problematic as RL agents are often difficult to interpret. 
Consideration of concepts like \textit{Explainability} \cite{10.1007/978-3-030-57321-8_5} is likely to be essential in providing the necessary administrative oversight, which may be necessary from both a risk and governance perspective. 
The most appropriate strategy is likely to involve an ensemble of methods, including collating a range of metrics that attest to the application's performance characteristics. 
These measures may include application-specific measures, like throughput and latency for a resource allocation application and include periodic utilization of AB testing to provide a relative measure against well-understood baselines.


In addition to the impact on reward acquisition, changes within the environment in which the RL agents exist may impact the computational performance of the model \cite{8258038}.
Metrics pertaining to model performance, like inference time, throughput, and RAM usage, will be important in identifying transient behaviours.

\subsection{Versioning in RLOps}
\textit{$*$ The challenges of versioning in RLOps lie in the synchronization management among the code, model, hyperparameters and developing tools.}

Versioning, or source control, is the practice of tracking and managing changes during development. O-RAN brings the opportunity to use software-based RICs with open interfaces widely. Flexible and fast iteration software development requires careful versioning, and this also applies to RLOps in O-RAN. 
\subsubsection{Data}
The data preparation in RL is different from SL or UL, as it comes from interacting with the environment. For communication network applications, data could come from either a running network or a DT. Live network data can be stored and versioned by data management tools like DVC~\footnote{https://dvc.org/}, Pachyderm~\footnote{https://www.pachyderm.com/} or other built-in tools in ML development frameworks. These tools attach version information to datasets. For artificial data generated by a DT, it is more efficient to give snapshots of the DT, including the simulation scenario, the configuration, the random seeds, etc. Given the versioning information of the DT, we should be able to reproduce the same dataset if needed.
\subsubsection{Model}
Versioning of the model is vital for controlling the model deployment, especially when facing environment changes or unexpected failures. Since the training pipeline of RL models for O-RAN takes both live network data and DT, it is important to version the training environment and pipeline as well as the model itself to trace back this self-learning approach. This includes the versioning of training configurations, the production environment, and the versioning of DT and network data mentioned in the previous section. The hyperparameters that correspond to each model should also be versioned.
\subsubsection{Code}
All the production code during development and deployment should be put into versioning. This includes the code to train the RL model, the code for testing and validation, the code for successfully deploying the trained model, and the application code. In addition, as the training of RL in O-RAN uses DT, the code for the DT development and deployment should also be versioned. The DT itself can be seen as a standalone project which requires proper source control~\cite{9374645}.

\subsection{Safety and Security in RLOps}
\textit{$*$ The challenges of safety and security in RLOps lie in the robustness assurance of the developed RL models.}

Model safety and operation security are critical for ML/RL applications in O-RAN. The former can be dealt with by introducing safety constraints into the \textit{Design} and \textit{Development} process. We discuss some principles to follow for the operation security, inspired by the DevSecOps~\cite{myrbakken2017devsecops}, which integrates security measures into the DevOps cycle.

For RL models running on wireless networks, \textit{Safety} is important for service assurance as well as avoiding catastrophic performance decay. In the exploratory learning phase, a common approach is to consider potential safety restrictions that exist in the environments, agents, and actions in advance and formalize them into a Constrained MDP (CMDP), which defines a constrained optimization problem as shown in equation \ref{eq:definition of CMDP}. A safety policy is expected to achieve by training on the CMDP \cite{dulac2019challenges}.
\begin{equation}
    \underset{\pi \in \Pi}{\mathrm{max}}\mathcal{G}(\pi) \quad s.t. \; C^k(\pi)\leq V_k, k =1,\cdots,K
    \label{eq:definition of CMDP}
\end{equation}
where $\mathcal{G}$ is the cumulative discounted reward of a policy $\pi$, $C^k(\pi)$ reflects the cumulative cost incurred by constraint $k$ on a given policy $\pi$. Specifically, $C^k$ can be defined as $c_k(s,a)$ which represents the possible constraint in terms of state $s$ and action $s$. \cite{dulac2019challenges} presents one solution to the CMDP, which is called Constrained Policy Optimization (CPO). It searches for the policy that maximizes the reward and satisfies the given constraints, i.e., safety requirements. In~\cite{hasanzadezonuzy2020learning_onlineCRL}, the sample efficiency in CMDP is further studied in a model-based manner. Robust MDP has also been considered in the scope of CMDP, leading to a robust soft-constrained solution to the Robust-CMDP problems~\cite{russel2020robustCMDP}.

\textit{Security} in communication networks protects the integrity of the system, including but not limited to data, applications and user privacy. The open interfaces in O-RAN bring democratised applications but also increase the chance for deployed applications to be attacked. Considering the potential fast and frequent developing circle enabled by the RLOps, security practices should be considered throughout the process. This is the emerging paradigm of DevSecOps, in which some of the security responsibility is downloaded to developers. In RLOps, we make several suggestions in addition to the standard DevSecOps.

Since RL is running in an interactive way to provide intelligent decisions to the system, it is essential to consider the feedback from the environment at the beginning, including the feedback on security. For example, a special state can be designed for the MDP to indicate the sudden change of agent behaviour, which could be a sign of attack. The adversarial agent can be introduced in the RL training to test the robustness against malicious agents~\cite{chen2019adversarial, elderman2017adversarial}. Inspired by~\cite{diaz2019selfmonitoring}, \textit{Monitoring} could also play an essential role in integrated security measures. Attack detection techniques like anomaly detection could be applied to enable security practices through monitoring.

\section{Best Practices of RLOps}
\label{sec:best practice}
In this section, we discuss some best practices and effective routines for successfully delivering RL applications as the reflections of general principles presented in Section~\ref{sec: principles RLOps}. We will elaborate on DT’s functionalities and critical features, and then discuss the automation and reproducibility engineering in RLOps, respectively.

\subsection{Digital Twins}
\label{sec:dt}
A wide range of working definitions of DT exists \cite{fuller2020digital}, where we consider the definition by \cite{erkoyuncu2018digital} which is that ``A digital twin is a digital representation of a physical item or assembly using integrated simulations and service data''. 
The standardization of wireless network DT is still in progress, but it should be able to provide high-fidelity representations of all components of the current live network.
This includes the RAN, core network, and characteristics of users and service behaviours among others.
Where each component will be modelled through the use of ML models or emulated elements, for example \cite{9374645}. 
As discussed in \cite{9374645}, DTs offer a wide range of benefits for communications networks, including reducing the deployment costs for new services and supporting network automation and optimization. 

Within the context of RLOps, DTs are likely to be an integral part of the development pipeline.
Enabling training, testing and validation of RL agents in an environment that provides a good approximation to the real world without the associated risks.
The key benefits it provides from an RL perspective are enumerated in the list below.
\begin{enumerate}
    \item \textbf{Exploration:} RL algorithms require exploration in order to learn about the environment in which they are operating. Exploration, by definition, is risky, as it requires the execution of actions that have potentially unknown outcomes and could, in principle, be unrecoverable \cite{DBLP:journals/corr/AmodeiOSCSM16}. A DT provides a high-fidelity approximation to the real network where a failure is an option, as any damage is inconsequential as it is reversible.  
    \item \textbf{Parallelization:} Sample efficiency is a crucial problem within RL, where agents typically take considerable time to train. The utilization of several environments in parallel can reduce the real clock time that an agent takes to converge \cite{horgan2018distributed,espeholt2018impala}. Deployment on the real network does not support this functionality.
    \item \textbf{Validation:} When any new component is added(be that physical hardware, an RL agent or some new software function), there is potential for unforeseen negative behaviours to occur. Mitigating these deployment risks is essential from a business perspective. A DT easily accommodates this desired functionality as it allows for simulation and investigation of network response in a wide variety of scenarios. From an RL-specific perspective, it allows for confirmation of the agent's capacity for reward acquisition and provides functionality to support the interpretability of RL policies more readily. 
\end{enumerate}

In addition to the number of compelling arguments for their utilization, certain risks must be realized, especially when the DT modelling can't reflect the networks' reality.
Within the remainder of this section, we introduce a well-known challenge considered by the RL robotics community, commonly referred to as \textit{Sim-to-Real} \cite{zhao2020sim}. 
The associated literature is concerned with training within simulation and deployment within the real world and attempts to mitigate risks associated with approximation error between the two systems. 
Fundamentally, this same desire and challenge will persist within our pipeline and more widely within telecommunications applications.
For a comprehensive survey of the area please refer to \cite{zhao2020sim}.

\begin{figure}[t]
    \centering
    \includegraphics[width=0.5\textwidth]{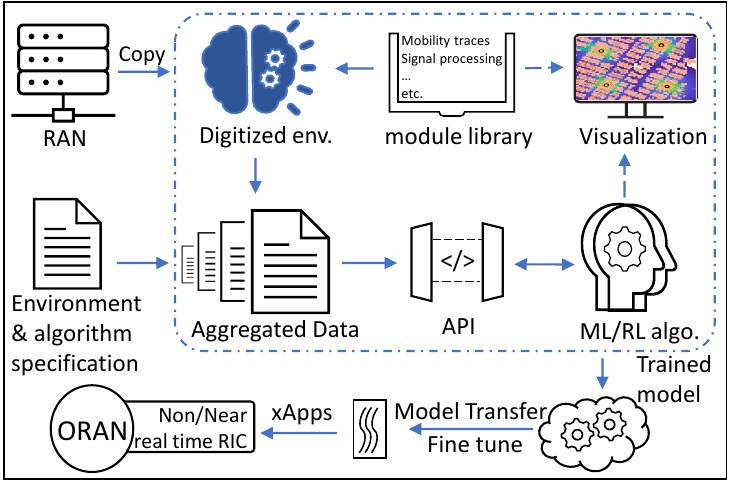}
    \caption{Digital twins functionalities. The core, which is a digital copy of the RAN, interacts with additional modules to simulate the change of network environments.}
    \label{fig:digital twin}
\end{figure}

\subsection{Automation}
The realization of an automated development process is undoubtedly the critical factor in any type of DevOps. The training procedure needs to be automated in order to save time and labour, expediting the transition from development to production.
\subsubsection{Data}
Data cleaning and preparation are often necessary for any new task or environment. This facilitates pattern detection for models as features are well scaled and ordered. As data generated for a task is often consistent, once the transformation procedure is done once, it can be repeated every other time without any need for manual interference. 
Following data preparation, appropriate feature/state representations must be created to be provided to the agent. This can include concatenating data frames from multiple time-steps together, skipping every $n$ frame, obtaining a certain embedding of the transformed data, etc. This process is often specific to the algorithm/task at hand and is done during training. Since it is a highly repeated step and requires no manual input past creation, it can be automated. The Data transformation and feature extraction process can be finished in the DML and PL layers of our network analytics platform, respectively.

Reward functions can be either extrinsically created for a problem or intrinsically generated from available data. The former case warrants no further automation; however, intrinsic reward signals are obtained from engineering pipelines that extract the signal out of the transformed data. This process will most likely be repeated on every training/evaluation step and must be automated. The data visualisation layer provides such information but needs to establish the automated reward engineering mechanism according to the specific cases.
\subsubsection{Model}
Given a certain environment or task, the data preparation pipeline, i.e., the network analytics platform can be triggered and completed automatically. This reduces the amount of time spent on data preparation and guarantees consistency as development evolves. Each RL model follows a specific training methodology. Following data preparation, the training process can also be automated. Training can terminate or resume given performance metrics attached to the agent. At last, for hyperparameter/parameter selection, a common process for DL can also be automated. A hyperparameter sweep can commence once the training pipeline is formulated. The best set of parameters can be chosen based on performance metrics.

\subsubsection{Code}
An evaluation/testing step can be automatically triggered once a model has completed training. If passed, the agent can then be deployed to production. This process requires creating rigorous testing scripts, bypassing the agent’s manual testing/evaluation, thereby automating the transition from development to production. A model is usually a small sub-part of a larger application infrastructure providing a specific service. Once a new agent is ready for production deployment, it is necessary to automate the new application build process to ensure each new version is well documented and tracked.

\subsection{Reproducibility}
In O-RAN, well-trained RL models may need to be widely deployed in a large geographic area. Therefore, in the face of different deployment environments and carriers, it is very important to ensure that the performance of the model does not deteriorate, that is the reproducibility.
\subsubsection{Data} The development cycle is often about the model, but in many cases can be about changing the environment or handling new data. Changes in data can break a model's performance, and retraining is usually necessary. Dealing with data changes without performance loss is of paramount importance in RL. An agent that can generalize is a flexible and robust one. To tackle the issue of generalization, research challenges have appeared in recent years, such as the Procgen challenge \cite{mohanty2021measuring} in which agents are tested on multiple versions of the same environment. Keeping track of older versions of data/environments is vital for maintaining stable versions, debugging drops in performance, and developing more robust models.

\subsubsection{Model} Model performance can change drastically with minor changes in the training algorithm. Reproducing results in RL is very difficult given its dynamic nature \cite{henderson2018deep}. In RL both the data source and the agent dynamically change. Moreover, they each influence one another. The environment affects how the agent trains, and the agent's policy impacts the environment's evolution. The ability to revert to stable versions of a model is vital for maintaining stability in the event of performance degradation. In terms of development, minor changes to the model can be researched on their own prior to compounding improvements. Maintaining a careful log of which models possess which mutations are important for ease of integration. Each model version should also contain its own pseudo-code, clearly elaborating the differences in the algorithm. Furthermore, the method of feature creation must be consistent and well logged as it affects how models interpret the provided data. Such strategies massively aid with the development and debugging of new models.

\subsubsection{Code} There will be specific dependencies upon which the model relies. Maintaining correct versioning between development and production is necessary for the replication of behaviour. The same goes for the software stack used to create the product in development. It makes no sense to rely on a different, untested stack in production. Therefore, it is often best to containerize development and production iterations. This means all versioning data is well documented within their own containers, allowing for ease of reproducibility.

\section{Proposed Data Analytics Platform for RLOps}
\label{sec:data engineering}
\begin{figure*}[t]
    \centering
    \includegraphics[width=1\textwidth]{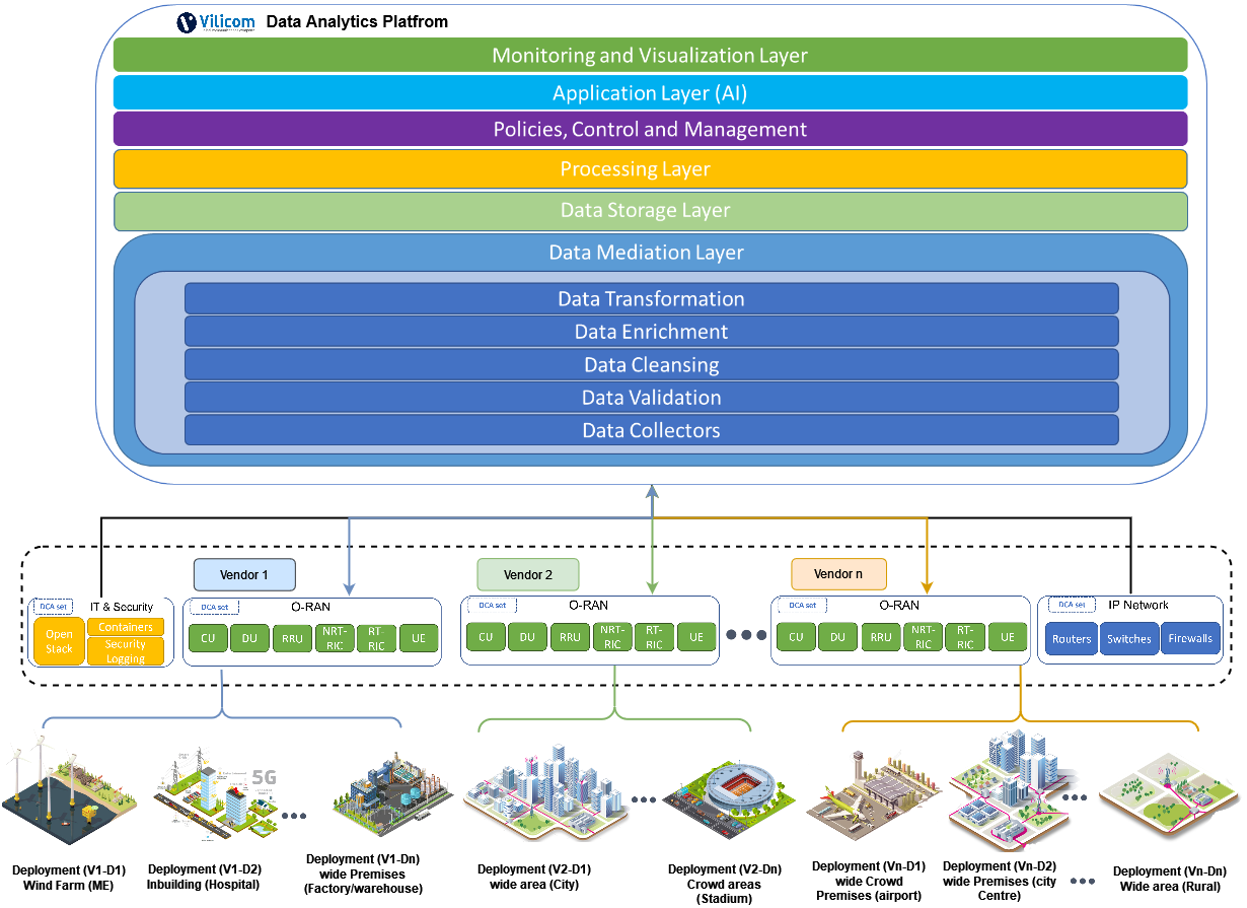}
    \caption{This figure illustrates the hierarchical definition and application scenarios of the proposed Network Analytics Platform.}
    \label{fig:platform}
\end{figure*}

The above sections illustrate the theoretical considerations of RLOps principles and best practices. In order to satisfy the above considerations and to effectively implement ML/RL on top of O-RAN interfaces,
a holistic data analytical platform rooting from RAN is necessary, which is helpful for DT continuous refinement, delivers the automation and reproducibility of RL models, and also fulfils the security and confidentiality of multi-tenancy public or private networks.  Hence, we design and implement the network analytics platform presented in Fig.~\ref{fig:platform}. We explain the compositions of this platform below.

\subsection{Features of the Data Analytics Platform}
\begin{itemize}
\item In this platform, the multiple raw data sources need to be collected, validated, enriched, transformed and stored in an integrated data pool. That needs to be processed by data engineering processes, such as application of business rules, creation of KPIs, feature engineering, linkage of data tables according to network topology mapping, etc., which ultimately enables the application of the algorithms according to the targeted use cases.
\item Besides, an O-RAN network is built on top of other system components such as IP networks and IT/Cloud infrastructures. The operation and maintenance of these systems are crucial for the overall network performance. It should be integrated into a holistic network management process that addresses all the components.
\item Since O-RAN is compliant and allows new architectural models based on multi-tenancy cloudified systems, the data pipeline must guarantee coherence and consistency in the treatment of the different data sources across the whole analytical cycle whilst maintaining strict compliance to the network segmentation and data confidentiality principles guaranteed by the interworking of the data storage, data processing and data governance and policy layers.
\end{itemize}

\subsection{Hierarchical Definition of the Data Analytics Platform}
\subsubsection{Data Collection Agents}
The data collection agents (DCA) are software applications deployed across the network layer, that interact with existing APIs and the network elements (NE). These agents use the standard APIs to collect the standard FCAPS dataset directly from NEs according to the use case. 
In a RAN, there are network domains that are implemented using equipment and technology that do not offer open and/or standard APIs. For that reason, it is necessary to develop a specific DCA designed to interact with the specific NE API or protocol, etc.
The DCA also has a function of data preparation right from the source, to allow for an efficient and effective data integration coming from multiple and diverse data sources, by normalising the data by applying the conventions that have been defined in the system. 
The DCA is also responsible for logging all its actions and performing initial data validation procedures. This function is important to trace end-to-end the data pipeline and assist the upper layer of the data mediation stack.
These applications are deployed directly on the NE's management plane or on adjacent servers. These have been designed to listen and track the data generated on these sources and can pull the logs and send them instantly to the data mediation layer (DML).
\subsubsection{Data Mediation Layer}
The DML is responsible for collecting the data by coordinating the DCAs in the southbound interface, data processing and implementing the northbound interface to the upper layers. This layer is a cluster-based system designed according to big data requirements and best practices \cite{8862913}, allowing the system to scale and support ultra-dense networks. After data is collected from the DCAs, the DML receives it in its raw format, requiring it to be prepared before going through validation and cleansing processes. The DML needs to add the schema information to the data stream and link it with the network topology. This preparation process increases the efficiency of the system by reducing the complexity of the data validation and data cleansing.

The DML is responsible for the data validation and data cleansing processes that consist in validating the data against the expected schema, identifying duplicate records, or missing records, and coping with latency on the data source in making the data records available.
It also prepares the dataset for an optimal application of the data enrichment processes that would fail if applied directly to the raw data due to missing network topology information. 
The data enrichment and transformation functions are tightly coupled with the data storage and processing layers because it prepares the data stream to match the schemas of the data lake and other consuming applications.
At the end of the DML cycle, the data offered to the upper layers are fully integrated, normalised, enriched and transformed according to the system conventions, thus simplifying the development of the data lake and of any processing applications. The DML layers can be continuously improved and extended to consume more – in quantity and diversity – data sources and to offer the data on the northbound interface in any format, type and frequency that is optimal to the layers consuming the data stream.
The DML coordinates with the DCAs to securely collect the data by implementing an encrypted data pipe. It creates one uniform data flow between each DCA and the upper layers.

\subsubsection{Data Storage Layer}
The data storage layer (DSL) contains one of the main components of the entire architecture which is the data lake. The data lake is the place where the data is stored to be made available to the upper layers, most importantly the processing and application layers. It is designed upon a scalable private cloud object storage; it provides the means to manage and store big datasets that come in diverse formats and structures and enables high throughput and fast access to the data.
The policies, business rules, network topology and other metadata required by the policies, control and management layer are stored in a dedicated relational database that is managed by the DSL.
Business Intelligence techniques and the development of ML/RL applications rely heavily upon wide and diverse historical datasets, for trend analysis, statistical analysis and for ML/RL in specific for model training, testing and validation. This demands many computational resources and requires DSL to be designed and implemented using big-data best practices \cite{249226}, to deliver optimal access to large-scale datasets.
On the other hand, feature engineering and RL-related tasks often require high-speed access to many disparate data sources to build and optimise the ML models; this requires high availability of some of the data in great quantities and diversity. For this, we have designed the data lake following the “Cold, Warm and Hot” approach \cite{dang2017graph}. 

The data lake is directly accessible by the other layers such as DM, processing and AI layer through a high throughput network. The design behind this storage system allows us to easily store petabytes of data and serve applications regardless of the data access requirements.

\subsubsection{Processing Layer}
The processing layer (PL) is composed of multiple applications deployed over a containerised environment that scales up with the increased demand from the services of the upper layers, such as the application and visualisation layers. 
The PL handles mainly three types of jobs, distributed real-time computation, distributed batch processing and jobs related to AI models such as environment states, reward calculation, AI model training/testing, etc.
AI and ML applications are complex and hard to develop, maintain, optimise, and deploy because of their iterative and multi-staged life-cycle. Complexity arises mostly from the stages that involve feature engineering, model training, model testing/validation and production deployment. On the other hand, the RL has more components to consider which are the environment, reward calculation, and the agents which make deploying these applications more challenging.
As emerged in MLOps practices, the main enhancement to solve the challenges of the AI lifecycle is to containerize all stages. The PL has been designed and implemented to follow this principle and overcome this challenge.
The PL allows the deployment and execution of services that underpin AI applications throughout its life cycle. In addition, to this, it also implements all the services that involve data processing, such as KPI calculation, real-time processing, alarm processing, online monitoring notifications, rule enforcement and data preparation for visualisation. This layer works in tandem with the lower layers, such as DSL and DML, to provide a containerised environment that simplifies the deployment and management of resource-intensive applications and guarantees high-throughput access to the data pool through dedicated and purpose-built data streams.
This layer will help to encapsulate the works in subphases where the task could be updated separately without affecting other phases. We illustrate some of the main jobs in this layer as follows:
\begin{itemize}
    \item KPI calculation: To measure the performance of the whole network, 3gpp produced a technical specification document for KPIs \cite{8862913}. These KPIs need an elevated level of domain expertise to develop and deploy across the data pool. The purpose of these KPIs includes but is not limited to the monitoring and troubleshooting of the network performance and long-term trend analysis of its performance. However, they are valuable features to build ML/RL models and reflect environment status. By abstracting this layer, we intend to save time and reduce complexity. The KPIs are calculated periodically. The results are eventually stored with the AI engineers' collected performance metrics for usage.
    \item Feature engineering and real-time data processing: Considering the requirements of the RLOps, the processing layer will also run applications that process streams and batches of data, so this layer is where the feature engineering process is done.
    \item ML/RL related components: The components needed to train, test, and validate the ML/RL application. These containers and the related applications are integrated into the whole platform so that they are able to cooperate with other containers and services offered in the processing layer. Additionally, the processing layer is able to run environment simulators or DT images and integrate them into the data pipeline.
\end{itemize}

\subsubsection{Policies and Control Layer}
The policies and control Layer is composed of a set of configuration methods, services and metadata that define and implement the business rules, object hierarchy and relationships that are relevant for the functionality implemented across the data mediation, data storage and processing layers.
The O-RAN FCAPS data, produced across the multiple virtual network functions (VNFs) and interfaces, is the most representative and important data type in this platform. This data is being structured and is not generated in its raw format, with the whole information that is required for its representation and integrated with other data sources. This layer contains the rules, metadata, and methodologies necessary for the efficient and effective implementation of the cycles of the DML and DSL, allowing the creation of the structures to validate, cleanse, enrich and store the data in an optimal format.
The network topology metadata and methods are fundamental for the linkage of the different managed objects and data structures, thus enabling the cross-layer analysis between network performance events and external events described by data sources - that are external to the O-RAN network and relevant to the analytical process, e.g. UE-based data that describes QoS and QoE events through detailed metrics and logs.
On the other hand, this layer also stores the policies and rules that control some aspects of the system’s cognitive capabilities, such as the identification of abnormal behaviour and respecting self-healing actions/decisions. These policies and rules can be defined by: subject matter experts (SMEs) through processes of data engineering, feature engineering and/or analytical engineering; and by automated analytic processes, possibly based on ML/RL applications that identify rules/decisions that after being validated and accepted by SMEs are later deployed on to production. 
\subsubsection{AI Layer (AI application management Layer)}
The AI layer is where the development, initial training and validation of the AI model happens. It allows the implementation of online training through real-time data consumption and offline model validation, generating results/decisions that are not implemented but rather validated by the developers and the subject matter experts. It also allows for monitoring logs and tracking the AI jobs' performance and related application images, mostly for testing and debugging purposes.
\subsubsection{Data Visualization Layer}
This layer is mostly dedicated to implementing business intelligence functions that allow SMEs to access the data in the format of graphical reports and dashboards, thus providing a visual interface to monitor the overall system performance.
Through this layer, it is possible to access reports and dashboards that inform about the performance of the different system components through the monitoring of dedicated measurements. 
The components that are monitored are: 
\begin{itemize}
    \item O-RAN network equipment, VNFs, protocols, interfaces, and functions: this allows for the network management SMEs to evaluate network performance, identify opportunities for optimisation, trends of systemic behaviour and evaluate the impact that AI algorithms might have on the overall system performance.
    \item AI application decision-making logging: this allows for the DevOps, MLOps and RLOps engineers to evaluate the performance of these applications during the entire life cycle from training to operations. It also allowed to report of the results of correlation and causation analysis visually and emphasised the evaluation of the decision of the application on the system performance.
\end{itemize}


\section{Conclusions}
\label{sec:Conclusion}
O-RAN embraces the intelligent models in specifications and treats ML/RL as a promising solution for achieving truly intelligent future network infrastructure. Considering the current lack of principles and practices for developing data-derived optimal decision-making strategies in O-RAN, we proposed the RLOps, which takes the life-cycle of RL model development as the main consideration, adopting the design, development, operations and safety/security as principles. We detail all main considerations and methodologies under these principles and integrate the above functions with the digital twins and the network analytics platform, which is geared to achieve automatic and reproducible model operations.

\section*{Acknowledgment}
This work was developed within the Innovate UK/CELTIC-NEXT European collaborative project on AIMM (AI-enabled Massive MIMO). This work has also been funded in part by the Next-Generation Converged Digital Infrastructure (NG-CDI) Project, supported by BT and Engineering and Physical Sciences Research Council (EPSRC), Grant ref. EP/R004935/1.

\bibliographystyle{IEEEtran} %
\bibliography{IEEEabrv,references} 

\begin{thebibliography}{10}
\providecommand{\url}[1]{#1}
\csname url@samestyle\endcsname
\providecommand{\newblock}{\relax}
\providecommand{\bibinfo}[2]{#2}
\providecommand{\BIBentrySTDinterwordspacing}{\spaceskip=0pt\relax}
\providecommand{\BIBentryALTinterwordstretchfactor}{4}
\providecommand{\BIBentryALTinterwordspacing}{\spaceskip=\fontdimen2\font plus
\BIBentryALTinterwordstretchfactor\fontdimen3\font minus
  \fontdimen4\font\relax}
\providecommand{\BIBforeignlanguage}[2]{{%
\expandafter\ifx\csname l@#1\endcsname\relax
\typeout{** WARNING: IEEEtran.bst: No hyphenation pattern has been}%
\typeout{** loaded for the language `#1'. Using the pattern for}%
\typeout{** the default language instead.}%
\else
\language=\csname l@#1\endcsname
\fi
#2}}
\providecommand{\BIBdecl}{\relax}
\BIBdecl

\bibitem{bonati2020open}
L.~Bonati, M.~Polese, S.~D’Oro, S.~Basagni, and T.~Melodia, ``Open,
  programmable, and virtualized 5{G} networks: State-of-the-art and the road
  ahead,'' \emph{Computer Networks}, vol. 182, p. 107516, 2020.

\bibitem{alliance2020ran}
O.~Alliance, ``O-ran use cases and deployment scenarios,'' \emph{White Paper,
  Feb}, 2020.

\bibitem{soltani2019deep}
M.~Soltani, V.~Pourahmadi, A.~Mirzaei, and H.~Sheikhzadeh, ``Deep
  learning-based channel estimation,'' \emph{IEEE Communications Letters},
  vol.~23, no.~4, pp. 652--655, 2019.

\bibitem{forster2007machine}
A.~Forster, ``Machine learning techniques applied to wireless ad-hoc networks:
  Guide and survey,'' in \emph{2007 3rd international conference on intelligent
  sensors, sensor networks and information}.\hskip 1em plus 0.5em minus
  0.4em\relax IEEE, 2007, pp. 365--370.

\bibitem{8493155}
X.~Chen, H.~Zhang, C.~Wu, S.~Mao, Y.~Ji, and M.~Bennis, ``Optimized computation
  offloading performance in virtual edge computing systems via deep
  reinforcement learning,'' \emph{IEEE Internet of Things Journal}, vol.~6,
  no.~3, pp. 4005--4018, 2019.

\bibitem{renggli2021data}
C.~Renggli, L.~Rimanic, N.~M. G{\"u}rel, B.~Karla{\v{s}}, W.~Wu, and C.~Zhang,
  ``A data quality-driven view of {MLOps},'' \emph{arXiv preprint
  arXiv:2102.07750}, 2021.

\bibitem{ebert2016devops}
C.~Ebert, G.~Gallardo, J.~Hernantes, and N.~Serrano, ``Devops,'' \emph{IEEE
  Software}, vol.~33, no.~3, pp. 94--100, 2016.

\bibitem{dunjko2018machine}
V.~Dunjko and H.~J. Briegel, ``Machine learning \& artificial intelligence in
  the quantum domain: {A} review of recent progress,'' \emph{Reports on
  Progress in Physics}, vol.~81, no.~7, p. 074001, 2018.

\bibitem{caron2018deep}
M.~Caron, P.~Bojanowski, A.~Joulin, and M.~Douze, ``Deep clustering for
  unsupervised learning of visual features,'' in \emph{Proceedings of the
  European Conference on Computer Vision (ECCV)}, 2018, pp. 132--149.

\bibitem{silverman2018density}
B.~W. Silverman, \emph{Density estimation for statistics and data
  analysis}.\hskip 1em plus 0.5em minus 0.4em\relax Routledge, 2018.

\bibitem{Bishop2006}
C.~Bishop, \emph{{Pattern Recognition and Machine Learning}}.\hskip 1em plus
  0.5em minus 0.4em\relax New York: Springer, 2006.

\bibitem{zhang2021multiagent}
K.~Zhang, Z.~Yang, and T.~Başar, ``Multi-agent reinforcement learning: A
  selective overview of theories and algorithms,'' 2021.

\bibitem{faisal_2021}
\BIBentryALTinterwordspacing
Faisal, ``Ran vs cloud ran vs vran vs o-ran: A simple guide!'' Apr 2021.
  [Online]. Available:
  \url{https://telcocloudbridge.com/blog/c-ran-vs-cloud-ran-vs-vran-vs-o-ran/}
\BIBentrySTDinterwordspacing

\bibitem{wypior2022open}
D.~Wypi{\'o}r, M.~Klinkowski, and I.~Michalski, ``Open ran—radio access
  network evolution, benefits and market trends,'' \emph{Applied Sciences},
  vol.~12, no.~1, p. 408, 2022.

\bibitem{akhtar2018threat}
N.~Akhtar and A.~Mian, ``Threat of adversarial attacks on deep learning in
  computer vision: A survey,'' \emph{IEEE Access}, vol.~6, pp.
  14\,410--14\,430, 2018.

\bibitem{otter2020survey}
D.~W. Otter, J.~R. Medina, and J.~K. Kalita, ``A survey of the usages of deep
  learning for natural language processing,'' \emph{IEEE Transactions on Neural
  Networks and Learning Systems}, vol.~32, no.~2, pp. 604--624, 2020.

\bibitem{wu2020reinforcement}
Y.-H. Wu, Z.-C. Yu, C.-Y. Li, M.-J. He, B.~Hua, and Z.-M. Chen, ``Reinforcement
  learning in dual-arm trajectory planning for a free-floating space robot,''
  \emph{Aerospace Science and Technology}, vol.~98, p. 105657, 2020.

\bibitem{silver2017mastering}
D.~Silver, J.~Schrittwieser, K.~Simonyan, I.~Antonoglou, A.~Huang, A.~Guez,
  T.~Hubert, L.~Baker, M.~Lai, A.~Bolton \emph{et~al.}, ``Mastering the game of
  go without human knowledge,'' \emph{nature}, vol. 550, no. 7676, pp.
  354--359, 2017.

\bibitem{yang2019aligraph}
H.~Yang, ``Aligraph: A comprehensive graph neural network platform,'' in
  \emph{Proceedings of the 25th ACM SIGKDD international conference on
  knowledge discovery \& data mining}, 2019, pp. 3165--3166.

\bibitem{jumper2021highly}
J.~Jumper, R.~Evans, A.~Pritzel, T.~Green, M.~Figurnov, O.~Ronneberger,
  K.~Tunyasuvunakool, R.~Bates, A.~{\v{Z}}{\'\i}dek, A.~Potapenko
  \emph{et~al.}, ``Highly accurate protein structure prediction with
  {AlphaFold},'' \emph{Nature}, vol. 596, no. 7873, pp. 583--589, 2021.

\bibitem{lien2011toward}
S.-Y. Lien, K.-C. Chen, and Y.~Lin, ``Toward ubiquitous massive accesses in
  {3GPP} machine-to-machine communications,'' \emph{IEEE Communications
  Magazine}, vol.~49, no.~4, pp. 66--74, 2011.

\bibitem{abedin2022elastic}
S.~F. Abedin, A.~Mahmood, N.~H. Tran, Z.~Han, and M.~Gidlund, ``Elastic {O-RAN}
  slicing for industrial monitoring and control: A distributed matching game
  and deep reinforcement learning approach,'' \emph{IEEE Transactions on
  Vehicular Technology}, 2022.

\bibitem{bonati2020intelligence}
L.~Bonati, S.~D'Oro, M.~Polese, S.~Basagni, and T.~Melodia, ``Intelligence and
  learning in {O-RAN} for data-driven {NextG} cellular networks,'' \emph{IEEE
  Communications Magazine}, vol.~59, no.~10, pp. 21--27, 2021.

\bibitem{filali2022communication}
A.~Filali, B.~Nour, S.~Cherkaoui, and A.~Kobbane, ``Communication and
  computation {O-RAN} resource slicing for {URLLC} services using deep
  reinforcement learning,'' \emph{arXiv preprint arXiv:2202.06439}, 2022.

\bibitem{9500721}
T.~Pamuklu, M.~Erol-Kantarci, and C.~Ersoy, ``Reinforcement learning based
  dynamic function splitting in disaggregated green {Open RANs},'' in \emph{ICC
  2021 - IEEE International Conference on Communications}, 2021, pp. 1--6.

\bibitem{9814869}
M.~Polese, L.~Bonati, S.~D’Oro, S.~Basagni, and T.~Melodia, ``Colo-ran:
  Developing machine learning-based {xApps} for {Open RAN} closed-loop control
  on programmable experimental platforms,'' \emph{IEEE Transactions on Mobile
  Computing}, pp. 1--14, 2022.

\bibitem{lien2021session}
S.-Y. Lien, D.-J. Deng, and B.-C. Chang, ``Session management for {URLLC} in
  {5G} open radio access network: A machine learning approach,'' in \emph{2021
  International Wireless Communications and Mobile Computing (IWCMC)}.\hskip
  1em plus 0.5em minus 0.4em\relax IEEE, 2021, pp. 2050--2055.

\bibitem{mollahasani2022energy}
S.~Mollahasani, T.~Pamuklu, R.~Wilson, and M.~Erol-Kantarci, ``Energy-aware
  dynamic {DU} selection and {NF} relocation in {O-RAN} using {Actor--Critic}
  learning,'' \emph{Sensors}, vol.~22, no.~13, p. 5029, 2022.

\bibitem{9679991}
O.~Orhan, V.~N. Swamy, T.~Tetzlaff, M.~Nassar, H.~Nikopour, and S.~Talwar,
  ``Connection management {xAPP} for {O-RAN RIC}: A graph neural network and
  reinforcement learning approach,'' in \emph{2021 20th IEEE International
  Conference on Machine Learning and Applications (ICMLA)}, 2021, pp. 936--941.

\bibitem{wang2021self}
X.~Wang, J.~D. Thomas, R.~J. Piechocki, S.~Kapoor, R.~Santos-Rodr{\'\i}guez,
  and A.~Parekh, ``Self-play learning strategies for resource assignment in
  {Open-RAN} networks,'' \emph{Computer Networks}, vol. 206, p. 108682, 2022.

\bibitem{iturria2022multi}
P.~E. Iturria-Rivera, H.~Zhang, H.~Zhou, S.~Mollahasani, and M.~Erol-Kantarci,
  ``Multi-agent team learning in virtualized open radio access networks
  {(O-RAN)},'' \emph{Sensors}, vol.~22, no.~14, p. 5375, 2022.

\bibitem{mungari2021rl}
F.~Mungari, ``An {RL} approach for radio resource management in the {O-RAN}
  architecture,'' in \emph{2021 18th Annual IEEE International Conference on
  Sensing, Communication, and Networking (SECON)}.\hskip 1em plus 0.5em minus
  0.4em\relax IEEE, 2021, pp. 1--2.

\bibitem{zhang2022team}
H.~Zhang, H.~Zhou, and M.~Erol-Kantarci, ``Team learning-based resource
  allocation for open radio access network {(O-RAN)},'' in \emph{ICC 2022 -
  IEEE International Conference on Communications}, 2022, pp. 4938--4943.

\bibitem{giannopoulos2022supporting}
A.~Giannopoulos, S.~Spantideas, N.~Kapsalis, P.~Gkonis, L.~Sarakis,
  C.~Capsalis, M.~Vecchio, and P.~Trakadas, ``Supporting intelligence in
  disaggregated open radio access networks: Architectural principles, {AI/ML}
  workflow, and use cases,'' \emph{IEEE Access}, vol.~10, pp. 39\,580--39\,595,
  2022.

\bibitem{bonati2022openran}
L.~Bonati, M.~Polese, S.~D’Oro, S.~Basagni, and T.~Melodia, ``{OpenRAN Gym}:
  An open toolbox for data collection and experimentation with {AI} in
  {O-RAN},'' in \emph{2022 IEEE Wireless Communications and Networking
  Conference (WCNC)}.\hskip 1em plus 0.5em minus 0.4em\relax IEEE, 2022, pp.
  518--523.

\bibitem{zhang2022federated}
H.~Zhang, H.~Zhou, and M.~Erol-Kantarci, ``Federated deep reinforcement
  learning for resource allocation in {O-RAN} slicing,'' \emph{arXiv preprint
  arXiv:2208.01736}, 2022.

\bibitem{ouyang2021next}
Y.~Ouyang, L.~Wang, A.~Yang, M.~Shah, D.~Belanger, T.~Gao, L.~Wei, and
  Y.~Zhang, ``The next decade of telecommunications artificial intelligence,''
  \emph{arXiv preprint arXiv:2101.09163}, 2021.

\bibitem{van2019deep}
T.~Van~Luong, Y.~Ko, N.~A. Vien, D.~H. Nguyen, and M.~Matthaiou, ``Deep
  learning-based detector for {OFDM-IM},'' \emph{IEEE wireless communications
  letters}, vol.~8, no.~4, pp. 1159--1162, 2019.

\bibitem{8445920}
A.~Felix, S.~Cammerer, S.~Dörner, J.~Hoydis, and S.~Ten~Brink,
  ``{OFDM-Autoencoder} for {End-to-End} learning of communications systems,''
  in \emph{2018 IEEE 19th International Workshop on Signal Processing Advances
  in Wireless Communications (SPAWC)}, 2018, pp. 1--5.

\bibitem{yang2019deep}
Y.~Yang, F.~Gao, G.~Y. Li, and M.~Jian, ``Deep learning-based downlink channel
  prediction for {FDD} massive {MIMO} system,'' \emph{IEEE Communications
  Letters}, vol.~23, no.~11, pp. 1994--1998, 2019.

\bibitem{wang2020precodernet}
Q.~Wang, K.~Feng, X.~Li, and S.~Jin, ``Precodernet: {Hybrid} beamforming for
  millimeter wave systems with deep reinforcement learning,'' \emph{IEEE
  Wireless Communications Letters}, vol.~9, no.~10, pp. 1677--1681, 2020.

\bibitem{wang2018unified}
Y.~Wang, Z.~Zhang, S.~Zhang, S.~Cao, and S.~Xu, ``A unified deep learning based
  polar-{LDPC} decoder for {5G} communication systems,'' in \emph{2018 10th
  International Conference on Wireless Communications and Signal Processing
  (WCSP)}.\hskip 1em plus 0.5em minus 0.4em\relax IEEE, 2018, pp. 1--6.

\bibitem{guo2022cooperative}
W.~Guo, N.~M.~F. Qureshi, I.~F. Siddiqui, and D.~R. Shin, ``Cooperative
  communication resource allocation strategies for {5G} and beyond networks: A
  review of architecture, challenges and opportunities,'' \emph{Journal of King
  Saud University-Computer and Information Sciences}, 2022.

\bibitem{bashir2019optimal}
A.~K. Bashir, R.~Arul, S.~Basheer, G.~Raja, R.~Jayaraman, and N.~M.~F. Qureshi,
  ``An optimal multitier resource allocation of cloud {RAN} in {5G} using
  machine learning,'' \emph{Transactions on emerging telecommunications
  technologies}, vol.~30, no.~8, p. e3627, 2019.

\bibitem{padakandla2021survey}
S.~Padakandla, ``A survey of reinforcement learning algorithms for dynamically
  varying environments,'' \emph{ACM Computing Surveys (CSUR)}, vol.~54, no.~6,
  pp. 1--25, 2021.

\bibitem{jordan2020evaluating}
S.~Jordan, Y.~Chandak, D.~Cohen, M.~Zhang, and P.~Thomas, ``Evaluating the
  performance of reinforcement learning algorithms,'' in \emph{International
  Conference on Machine Learning}.\hskip 1em plus 0.5em minus 0.4em\relax PMLR,
  2020, pp. 4962--4973.

\bibitem{arulkumaran2017deep}
K.~Arulkumaran, M.~P. Deisenroth, M.~Brundage, and A.~A. Bharath, ``Deep
  reinforcement learning: A brief survey,'' \emph{IEEE Signal Processing
  Magazine}, vol.~34, no.~6, pp. 26--38, 2017.

\bibitem{moerland2020model}
T.~M. Moerland, J.~Broekens, and C.~M. Jonker, ``Model-based reinforcement
  learning: A survey,'' \emph{arXiv preprint arXiv:2006.16712}, 2020.

\bibitem{puiutta2020explainable}
E.~Puiutta and E.~M. Veith, ``Explainable reinforcement learning: A survey,''
  in \emph{International Cross-Domain Conference for Machine Learning and
  Knowledge Extraction}.\hskip 1em plus 0.5em minus 0.4em\relax Springer, 2020,
  pp. 77--95.

\bibitem{fuller2020digital}
A.~Fuller, Z.~Fan, C.~Day, and C.~Barlow, ``Digital twin: Enabling
  technologies, challenges and open research,'' \emph{IEEE Access}, vol.~8, pp.
  108\,952--108\,971, 2020.

\bibitem{9374645}
H.~X. Nguyen, R.~Trestian, D.~To, and M.~Tatipamula, ``Digital twin for {5G}
  and beyond,'' \emph{IEEE Communications Magazine}, vol.~59, no.~2, pp.
  10--15, 2021.

\bibitem{zhao2020sim}
W.~Zhao, J.~P. Queralta, and T.~Westerlund, ``Sim-to-real transfer in deep
  reinforcement learning for robotics: {A} survey,'' in \emph{2020 IEEE
  Symposium Series on Computational Intelligence (SSCI)}.\hskip 1em plus 0.5em
  minus 0.4em\relax IEEE, 2020, pp. 737--744.

\bibitem{9158349}
A.~Kadian, J.~Truong, A.~Gokaslan, A.~Clegg, E.~Wijmans, S.~Lee, M.~Savva,
  S.~Chernova, and D.~Batra, ``{Sim2Real Predictivity}: Does evaluation in
  simulation predict real-world performance?'' \emph{IEEE Robotics and
  Automation Letters}, vol.~5, no.~4, pp. 6670--6677, 2020.

\bibitem{springenberg2016bayesian}
J.~T. Springenberg, A.~Klein, S.~Falkner, and F.~Hutter, ``Bayesian
  optimization with robust bayesian neural networks,'' \emph{Advances in neural
  information processing systems}, vol.~29, pp. 4134--4142, 2016.

\bibitem{li2017deep}
Y.~Li, ``Deep reinforcement learning: An overview,'' \emph{arXiv preprint
  arXiv:1701.07274}, 2017.

\bibitem{gauci2018horizon}
J.~Gauci, E.~Conti, Y.~Liang, K.~Virochsiri, Y.~He, Z.~Kaden, V.~Narayanan,
  X.~Ye, Z.~Chen, and S.~Fujimoto, ``Horizon: Facebook's open source applied
  reinforcement learning platform,'' \emph{arXiv preprint arXiv:1811.00260},
  2018.

\bibitem{DBLP:journals/corr/abs-2004-05785}
\BIBentryALTinterwordspacing
J.~Lu, A.~Liu, F.~Dong, F.~Gu, J.~Gama, and G.~Zhang, ``Learning under concept
  drift: {A} review,'' \emph{CoRR}, vol. abs/2004.05785, 2020. [Online].
  Available: \url{https://arxiv.org/abs/2004.05785}
\BIBentrySTDinterwordspacing

\bibitem{medel2016modelling}
V.~Medel, O.~Rana, J.~{\'A}. Ba{\~n}ares, and U.~Arronategui, ``Modelling
  performance \& resource management in kubernetes,'' in \emph{Proceedings of
  the 9th International Conference on Utility and Cloud Computing}, 2016, pp.
  257--262.

\bibitem{alwarafy2021deep}
A.~Alwarafy, M.~Abdallah, B.~S. Ciftler, A.~Al-Fuqaha, and M.~Hamdi, ``Deep
  reinforcement learning for radio resource allocation and management in next
  generation heterogeneous wireless networks: A survey,'' \emph{arXiv preprint
  arXiv:2106.00574}, 2021.

\bibitem{altman1999constrained}
E.~Altman, \emph{Constrained Markov decision processes}.\hskip 1em plus 0.5em
  minus 0.4em\relax CRC Press, 1999, vol.~7.

\bibitem{myrbakken2017devsecops}
H.~Myrbakken and R.~Colomo-Palacios, ``Devsecops: {A} multivocal literature
  review,'' in \emph{International Conference on Software Process Improvement
  and Capability Determination}.\hskip 1em plus 0.5em minus 0.4em\relax
  Springer, 2017, pp. 17--29.

\bibitem{chen2019adversarial}
T.~Chen, J.~Liu, Y.~Xiang, W.~Niu, E.~Tong, and Z.~Han, ``Adversarial attack
  and defense in reinforcement learning-from {AI} security view,''
  \emph{Cybersecurity}, vol.~2, no.~1, pp. 1--22, 2019.

\bibitem{elderman2017adversarial}
R.~Elderman, L.~J. Pater, A.~S. Thie, M.~M. Drugan, and M.~A. Wiering,
  ``Adversarial reinforcement learning in a cyber security simulation.'' in
  \emph{ICAART (2)}, 2017, pp. 559--566.

\bibitem{pattanaik2017robust}
A.~Pattanaik, Z.~Tang, S.~Liu, G.~Bommannan, and G.~Chowdhary, ``Robust deep
  reinforcement learning with adversarial attacks,'' \emph{arXiv preprint
  arXiv:1712.03632}, 2017.

\bibitem{brockman2016openai}
G.~Brockman, V.~Cheung, L.~Pettersson, J.~Schneider, J.~Schulman, J.~Tang, and
  W.~Zaremba, ``Openai gym,'' \emph{arXiv preprint arXiv:1606.01540}, 2016.

\bibitem{dulac2019challenges}
G.~Dulac-Arnold, N.~Levine, D.~J. Mankowitz, J.~Li, C.~Paduraru, S.~Gowal, and
  T.~Hester, ``Challenges of real-world reinforcement learning: definitions,
  benchmarks and analysis,'' \emph{Machine Learning}, vol. 110, no.~9, pp.
  2419--2468, 2021.

\bibitem{43146}
D.~Sculley, G.~Holt, D.~Golovin, E.~Davydov, T.~Phillips, D.~Ebner,
  V.~Chaudhary, and M.~Young, ``Machine learning: The high interest credit card
  of technical debt,'' in \emph{SE4ML: Software Engineering for Machine
  Learning (NIPS 2014 Workshop)}, 2014.

\bibitem{levine2020offlineRL}
S.~Levine, A.~Kumar, G.~Tucker, and J.~Fu, ``Offline reinforcement learning:
  Tutorial, review, and perspectives on open problems,'' \emph{arXiv preprint
  arXiv:2005.01643}, 2020.

\bibitem{8258038}
E.~Breck, S.~Cai, E.~Nielsen, M.~Salib, and D.~Sculley, ``The {ML} test score:
  A rubric for {ML} production readiness and technical debt reduction,'' in
  \emph{2017 IEEE International Conference on Big Data (Big Data)}, 2017, pp.
  1123--1132.

\bibitem{9185035}
D.~Guo, L.~Tang, X.~Zhang, and Y.-C. Liang, ``Joint optimization of handover
  control and power allocation based on multi-agent deep reinforcement
  learning,'' \emph{IEEE Transactions on Vehicular Technology}, vol.~69,
  no.~11, pp. 13\,124--13\,138, 2020.

\bibitem{shapley1953stochastic}
L.~S. Shapley, ``Stochastic games,'' \emph{Proceedings of the national academy
  of sciences}, vol.~39, no.~10, pp. 1095--1100, 1953.

\bibitem{florensa2017stochastic}
\BIBentryALTinterwordspacing
C.~Florensa, Y.~Duan, and P.~Abbeel, ``Stochastic neural networks for
  hierarchical reinforcement learning,'' in \emph{International Conference on
  Learning Representations}, 2017. [Online]. Available:
  \url{https://openreview.net/forum?id=B1oK8aoxe}
\BIBentrySTDinterwordspacing

\bibitem{rao2020make}
N.~Rao, E.~Aljalbout, A.~Sauer, and S.~Haddadin, ``How to make deep {RL} work
  in practice,'' 2020.

\bibitem{mnih2013playing}
V.~Mnih, K.~Kavukcuoglu, D.~Silver, A.~Graves, I.~Antonoglou, D.~Wierstra, and
  M.~Riedmiller, ``Playing {Atari} with deep reinforcement learning,'' 2013.

\bibitem{hochreiter1997long}
S.~Hochreiter and J.~Schmidhuber, ``Long short-term memory,'' \emph{Neural
  computation}, vol.~9, no.~8, pp. 1735--1780, 1997.

\bibitem{sutton2018reinforcement}
R.~S. Sutton and A.~G. Barto, \emph{Reinforcement learning: An
  introduction}.\hskip 1em plus 0.5em minus 0.4em\relax MIT press, 2018.

\bibitem{narvekar2018learning}
S.~Narvekar and P.~Stone, ``Learning curriculum policies for reinforcement
  learning,'' in \emph{Proceedings of the 18th International Conference on
  Autonomous Agents and MultiAgent Systems}, ser. AAMAS '19.\hskip 1em plus
  0.5em minus 0.4em\relax Richland, SC: International Foundation for Autonomous
  Agents and Multiagent Systems, 2019, p. 25–33.

\bibitem{hussein2017imitation}
A.~Hussein, M.~M. Gaber, E.~Elyan, and C.~Jayne, ``Imitation learning: A survey
  of learning methods,'' \emph{ACM Computing Surveys (CSUR)}, vol.~50, no.~2,
  pp. 1--35, 2017.

\bibitem{valencia2019using}
R.~Valencia, C.-W. Sham, and O.~Sinnen, ``Using neuroevolved binary neural
  networks to solve reinforcement learning environments,'' in \emph{2019 IEEE
  Asia Pacific Conference on Circuits and Systems (APCCAS)}.\hskip 1em plus
  0.5em minus 0.4em\relax IEEE, 2019, pp. 301--304.

\bibitem{ghavamzadeh2016bayesian}
M.~Ghavamzadeh, S.~Mannor, J.~Pineau, A.~Tamar \emph{et~al.}, ``Bayesian
  reinforcement learning: A survey,'' \emph{Foundations and
  Trends{\textregistered} in Machine Learning}, vol.~8, no. 5-6, pp. 359--483,
  2015.

\bibitem{strens2000bayesian}
M.~Strens, ``A bayesian framework for reinforcement learning,'' in \emph{ICML},
  vol. 2000, 2000, pp. 943--950.

\bibitem{google}
\BIBentryALTinterwordspacing
``{MLOps}: Continuous delivery and automation pipelines in machine learning.''
  [Online]. Available:
  \url{https://cloud.google.com/architecture/mlops-continuous-delivery-and-automation-pipelines-in-machine-learning}
\BIBentrySTDinterwordspacing

\bibitem{DBLP:journals/corr/abs-1906-04737}
\BIBentryALTinterwordspacing
G.~Papoudakis, F.~Christianos, A.~Rahman, and S.~V. Albrecht, ``Dealing with
  non-stationarity in multi-agent deep reinforcement learning,'' \emph{CoRR},
  vol. abs/1906.04737, 2019. [Online]. Available:
  \url{http://arxiv.org/abs/1906.04737}
\BIBentrySTDinterwordspacing

\bibitem{DBLP:journals/corr/AmodeiOSCSM16}
\BIBentryALTinterwordspacing
D.~Amodei, C.~Olah, J.~Steinhardt, P.~F. Christiano, J.~Schulman, and
  D.~Man{\'{e}}, ``Concrete problems in {AI} safety,'' \emph{CoRR}, vol.
  abs/1606.06565, 2016. [Online]. Available:
  \url{http://arxiv.org/abs/1606.06565}
\BIBentrySTDinterwordspacing

\bibitem{10.1007/978-3-030-57321-8_5}
E.~Puiutta and E.~M. S.~P. Veith, ``{Explainable Reinforcement Learning: A
  Survey},'' in \emph{Machine Learning and Knowledge Extraction}, A.~Holzinger,
  P.~Kieseberg, A.~M. Tjoa, and E.~Weippl, Eds.\hskip 1em plus 0.5em minus
  0.4em\relax Cham: Springer International Publishing, 2020, pp. 77--95.

\bibitem{hasanzadezonuzy2020learning_onlineCRL}
A.~HasanzadeZonuzy, A.~Bura, D.~Kalathil, and S.~Shakkottai, ``Learning with
  safety constraints: Sample complexity of reinforcement learning for
  constrained mdps,'' in \emph{Proceedings of the AAAI Conference on Artificial
  Intelligence}, vol.~35, no.~9, 2021, pp. 7667--7674.

\bibitem{russel2020robustCMDP}
R.~Hasan~Russel, M.~Benosman, and J.~Van~Baar, ``Robust constrained-mdps:
  Soft-constrained robust policy optimization under model uncertainty,''
  \emph{arXiv e-prints}, pp. arXiv--2010, 2020.

\bibitem{diaz2019selfmonitoring}
J.~D{\'\i}az, J.~E. P{\'e}rez, M.~A. Lopez-Pe{\~n}a, G.~A. Mena, and
  A.~Yag{\"u}e, ``Self-service cybersecurity monitoring as enabler for
  devsecops,'' \emph{IEEE Access}, vol.~7, pp. 100\,283--100\,295, 2019.

\bibitem{erkoyuncu2018digital}
J.~A. Erkoyuncu, P.~Butala, R.~Roy \emph{et~al.}, ``Digital twins:
  Understanding the added value of integrated models for through-life
  engineering services,'' \emph{Procedia Manufacturing}, vol.~16, pp. 139--146,
  2018.

\bibitem{horgan2018distributed}
\BIBentryALTinterwordspacing
D.~Horgan, J.~Quan, D.~Budden, G.~Barth-Maron, M.~Hessel, H.~van Hasselt, and
  D.~Silver, ``Distributed prioritized experience replay,'' in
  \emph{International Conference on Learning Representations}, 2018. [Online].
  Available: \url{https://openreview.net/forum?id=H1Dy---0Z}
\BIBentrySTDinterwordspacing

\bibitem{espeholt2018impala}
L.~Espeholt, H.~Soyer, R.~Munos, K.~Simonyan, V.~Mnih, T.~Ward, Y.~Doron,
  V.~Firoiu, T.~Harley, I.~Dunning, S.~Legg, and K.~Kavukcuoglu, ``{IMPALA:
  Scalable Distributed Deep-RL with Importance Weighted Actor-Learner
  Architectures},'' 2018.

\bibitem{mohanty2021measuring}
\BIBentryALTinterwordspacing
S.~Mohanty, J.~Poonganam, A.~Gaidon, A.~Kolobov, B.~Wulfe, D.~Chakraborty,
  G.~\u{S}emetulskis, J.~a. Schapke, J.~Kubilius, J.~Pa\"ukonis, L.~Klimas,
  M.~Hausknecht, P.~MacAlpine, Q.~N. Tran, T.~Tumiel, X.~Tang, X.~Chen,
  C.~Hesse, J.~Hilton, W.~H. Guss, S.~Genc, J.~Schulman, and K.~Cobbe,
  ``Measuring sample efficiency and generalization in reinforcement learning
  benchmarks: Neurips 2020 procgen benchmark,'' in \emph{Proceedings of the
  NeurIPS 2020 Competition and Demonstration Track}, ser. Proceedings of
  Machine Learning Research, H.~J. Escalante and K.~Hofmann, Eds., vol.
  133.\hskip 1em plus 0.5em minus 0.4em\relax PMLR, 06--12 Dec 2021, pp.
  361--395. [Online]. Available:
  \url{https://proceedings.mlr.press/v133/mohanty21a.html}
\BIBentrySTDinterwordspacing

\bibitem{henderson2018deep}
P.~Henderson, R.~Islam, P.~Bachman, J.~Pineau, D.~Precup, and D.~Meger, ``Deep
  reinforcement learning that matters,'' in \emph{Proceedings of the AAAI
  conference on artificial intelligence}, vol.~32, no.~1, 2018.

\bibitem{8862913}
Y.~Roh, G.~Heo, and S.~E. Whang, ``A survey on data collection for machine
  learning: A big data - {AI} integration perspective,'' \emph{IEEE
  Transactions on Knowledge and Data Engineering}, vol.~33, no.~4, pp.
  1328--1347, 2021.

\bibitem{249226}
W.~Chang and N.~Grady, ``\BIBforeignlanguage{en}{Nist big data interoperability
  framework: Volume 1, definitions},'' 2019-10-21 2019.

\bibitem{dang2017graph}
T.-H. Dang-Ha, D.~Roverso, and R.~Olsson, ``Graph of virtual actors ({GOVA}): A
  big data analytics architecture for {IOT},'' in \emph{2017 IEEE International
  Conference on Big Data and Smart Computing (BigComp)}.\hskip 1em plus 0.5em
  minus 0.4em\relax IEEE, 2017, pp. 162--169.

\end{thebibliography}

\end{document}